\documentclass[usegraphicx]{mn2e}

 \usepackage{amssymb}

\newif\ifAMStwofonts

\newcommand{\Halpha}{H$\rm \alpha$}

\newcommand{\degrees}{$\rm ^{\circ}$}

\begin{document}

\title[Stellar populations in the 6dFGS
]
{The effects of stellar populations on galaxy scaling relations 
in the 6dF Galaxy Survey
}
\author[
Proctor et al. 
]
{
Robert N. Proctor$^{1}$,  Philip Lah$^{2}$, Duncan A. Forbes$^{1}$, Matthew
Colless$^{3}$\\ 
\\
\LARGE and Warrick Couch$^{1}$ \\ 
$^1$ Centre for Astrophysics \& Supercomputing, Swinburne University,
Hawthorn VIC 3122, Australia\\
Email: rproctor@astro.swin.edu.au, dforbes@astro.swin.edu.au, wcouch@astro.swin.edu.au \\
$^2$ Research School of Astronomy \& Astrophysics, Australian National
University, Weston Creek, ACT 2611, Australia \\
Email: plah@mso.anu.edu.au\\
$^3$ Anglo-Australian Observatory, PO Box 296, Epping, NSW 1710, Australia\\
Email: colless@aao.gov.au}

\pagerange{\pageref{firstpage}--\pageref{lastpage}}
\def\LaTeX{L\kern-.36em\raise.3ex\hbox{a}\kern-.15em
    T\kern-.1667em\lower.7ex\hbox{E}\kern-.125emX}

\newtheorem{theorem}{Theorem}[section]

\label{firstpage}

\newpage

\maketitle

\begin{abstract}

We present an analysis of the stellar populations as a function of
mass in a sample of $\sim$7000 galaxies of all morphological and
emission types from the 6dF Galaxy Survey (6dFGS). We measure velocity
dispersions and Lick indices from the spectra of the central regions
of these galaxies, deriving ages and metallicities from the Lick
indices using stellar population models. We also derive dynamical masses and
dynamical mass-to-light ratios for these galaxies by combining the central
velocity dispersions with global photometry in the B, R and K bands
from SuperCOSMOS and 2MASS. Together, these data allow us to reduce
the degeneracies between age, metallicity and star formation 
burst-strength that have limited previous studies.

We find that old galaxies exhibit a mass-metallicity relation with
logarithmic slope d[Fe/H]/dlog$M \approx 0.25$, while young galaxies
show slopes consistent with zero. When we account for the effects of
the mass-metallicity relation, we obtain a single, consistent relation
between mass-to-light ratio and mass for old galaxies in all passbands,
$M/L \propto M^{0.15}$. As we have accounted for stellar population
effects, this remaining variation in the mass-to-light with mass (the
residual `tilt' of the Fundamental Plane) must have a dynamical
origin.  However, we demonstrate that any simple trend between
mass-to-light-ratio and mass or luminosity is inconsistent with the
observations, and that a more complex relationship must exist.

We find that the central regions of galaxies of all masses often
exhibit young stellar populations. However it is only in the
lowest-mass galaxies studied ($\sim10^{10}\,M_{\odot}$) that these
populations are evident in the global photometry. In higher-mass
galaxies, young central populations have decreasing influence on the
global photometry, with there being no discernible impact in galaxies 
more massive than $\sim2\times10^{11}\,M_{\odot}$. 
We conclude that the young stellar populations detected in spectroscopic
studies are generally centrally concentrated, and that there is an
upper limit on the mass of star-forming events in massive galaxies.
These results have important ramifications for mass-to-light ratios 
estimated from photometric observations.

\end{abstract}

\begin{keywords}
galaxies: general, galaxies: stellar content, galaxies: kinematics and dynamics 

\end{keywords}

\section{Introduction}
Current models of galaxy formation and assembly can be characterised
by two extreme and competing views. One such view is represented by
the early "monolithic collapse" model, whereby the spheroidal (bulge)
components of galaxies form by the early collapse of individual gas
clouds (e.g. Larson 1974; Carlberg 1984). In these early collapse
models the star formation induced by the collapse is terminated when
the source of gas is either consumed or expelled from the galaxy. The
bulge then ages passively (albeit perhaps subject to the later
addition of a disk). Such models are largely motivated by the small
scatter in the observed relationships between colour and magnitude
(e.g. Baum 1959; Bower, Lucy \& Ellis 1992a,b) and velocity
dispersion, surface brightness and size (the Fundamental Plane;
Dressler et al. 1987) for early-type galaxies and bulges. To explain
the tightness of these relations, early collapse models require that
star formation in bulges terminates at early times (z$\ge$2; Bower,
Lucy \& Ellis 1992a,b), so that bulge dominated (e.g. elliptical)
galaxies should be uniformly old.

This picture faces challenges on a number of fronts.  For instance,
the differing slopes found for the Fundamental Plane in various bands
(e.g. Bernardi et al. 2003b) are difficult to explain in the monolithic
collapse scenario. Furthermore, spectroscopic studies of the stellar
populations in the bulges of galaxies find central ages ranging from
$\sim$1--15 Gyr (e.g. Trager et al. 2000; Proctor \&
Sansom 2002; Terlevich \& Forbes 2002: Caldwell, Rose \& Concannon
2003; Kauffmann et al. 2003a,b; Proctor et al.  2004b), indicating
evidence for more recent star formation.

However, by far the most important challenge to the early collapse
picture is the fundamental differences in galaxy formation and
assembly histories that are predicted when the growth of structure via
gravitational instability is modelled for a dark matter-dominated
universe, now characterised extremely well via precision measurements
of the cosmological parameters (e.g. the $Lambda$CDM model; Spergel et
al. 2003). Such "hierarchical" models, which represent the other
extreme view, predict galaxy-mass objects to be assembled by the
successive merging of lower-mass objects via hierarchical merging
(White \& Rees 1978), usually over an extended period.

Another important prediction of hierarchical merging models is that
the merger rate of galaxies is a function of environment, i.e. on
average, galaxies in the dense environs of clusters merge early, while
those in the less dense `field' merge later (Kauffmann 1996). In the
hierarchical merging scenario the properties of galaxies are therefore
subject to strong influences from their environment.  Results from
stellar population studies appear to confirm the prediction of
hierarchical merging models in that `field' galaxies possess slightly
younger central ages than their cluster counterparts (e.g. Thomas et
al., 2005; Bernardi et al., 2006; Smith et al., 2006).

In contrast to early collapse models, hierarchical merging models
predict galaxy bulges to possess a range of ages, with low-mass
galaxies forming earlier, and on shorter timescales, than high-mass
galaxies.  At first glance, the wide range of ages observed in
spectroscopic studies might also appear to confirm the hierarchical
merging predictions, however the observed trend of age with mass is
for lower-mass systems to be {\it younger} than higher-mass
systems. This is inconsistent with hierarchical merging models which
predict lower-mass systems to form earlier than high-mass system
(Kauffmann 1996). The observed trend is therefore sometimes referred
to as `anti-hierarchical'.

The recent De Lucia et al. (2006) analysis of the Millennium
Simulation provides insights important to the
development of a coherent picture of galaxy evolution. The Millennium
Simulation models the evolution of dark matter using an N-body code to
which De Lucia et al.  couple a semi-analytic model of galaxy
formation. De Lucia et al's analysis clearly demonstrates that a
strong distinction must be made between the {\it mass assembly} of galaxies
and the {\it formation} of their stellar populations. This can be
understood intuitively by the simple realisation that mergers bring
old stars into a merger remnant as well as making new stars. Merger
remnants may therefore be, and often are, still dominated (in terms of
their mass) by old stellar populations.  De Lucia et al. conclude that
while the mass assembly of galaxies is indeed hierarchical, their star
formation histories are nevertheless anti-hierarchical. Despite this
improvement in our theoretical understanding, the problem remains
how to reconcile the broad range of predicted (and spectroscopically
observed) galaxy ages with the extremely small scatter in some scaling
relations.

Recent surveys of large numbers of nearby galaxies have provided
statistically significant results and hence further observational
constraints on galaxy formation models. Such surveys include the 2dF
Galaxy Redshift Survey (2dFGRS: Colless et al. 2001b) and the Sloan
Digital Sky Survey (SDSS; York et al. 2000).  From central region
spectra and global optical colours the 2dFGRS probed the cosmic star
formation history at relatively low redshift (Baldry et al. 2002) and
its variation with local environmental density (Lewis et al. 2002).
The SDSS also obtained optical colours and central spectra for a large
number of low redshift galaxies. This ongoing dataset has been used by
several authors to study the star formation history and scaling
relations of nearby galaxies. These studies include Bernardi et
al. (2003b) who investigated scaling relations for $\sim$9,000
galaxies, Eisenstein et al. (2003) who combined spectra of 22,000
massive galaxies, Gallazzi et al. (2005) who probed the star formation
history of $\sim$40,000 galaxies, Cid Fernandes et al. (2005) who
conducted a spectral synthesis of $\sim$50,000 galaxies, and Chang et
al. (2006) who combined spectra and optical colours with 2MASS
near-infrared colours for $\sim$3,000 galaxies. Such large datasets
have allowed these authors to investigate trends as a function of
mass, environment, galaxy type etc. Already, a galaxy mass-metallicity
relation has been detected in the metallicities of both gas (Tremonti
et al. 2004) and stars (Gallazzi et al. 2005) and strong trends with
galaxy environment identified (Lewis et al. 2002).

However, the above studies have not explicitly explored the link
between the {\it central} stellar populations, which they observe, and
the {\it global} photometry of the galaxies in their samples.  It is
this issue that this paper specifically addresses.  To this end, we compare
the results of stellar population analysis of spectra from the 6dF
Galaxy Survey (6dFGS; Jones et al. 2004) to the global near-infrared
photometry of the 2MASS Extended Source Catalog (2MASS XSC; Jarrett
et al. 2000a,b) and optical SuperCOSMOS (Hambly, Irwin \&
MacGillivray. 2001) data for several thousand galaxies. We explore
such issues as the fraction of galaxies' stellar populations that can
have been formed in recent star formation events and the effects of
stellar populations on the slopes of colour-magnitude relations and
the varying `tilt' of the Fundamental Plane.

The paper is laid out as follows. In the next section we outline the 
photometric and spectroscopic data. In Section \ref{spectroscopicanal} 
we detail the spectroscopic data reductions and analysis used to derive stellar
population parameters. Section \ref{results} presents the results of our 
analysis, which are discussed and summarised in Section \ref{concs}.

\section{The sample data}
\label{sampledata}

\subsection{The 6\lowercase{d}F Galaxy Survey}
The 6dFGS is a spectroscopic survey of the entire southern 
sky more than 10 degrees from the Galactic Plane. The survey measured 
redshifts for more than 124,000 galaxies.  The primary survey targets 
were selected primarily from the 2MASS Extended Source Catalog (XSC). 
All galaxies brighter than K$_{tot}$=12.75 were included in the sample. 
Secondary samples of 2MASS and SuperCOSMOS galaxies,
complete down to magnitude limits 
(H,~J,~r$_F$,~b$_J$)~=~(13.00,~13.75,~15.60,~16.75), were also included.  

The survey was carried out on the UK Schmidt Telescope using the
Six-Degree Field (6dF) multi-fibre spectrograph. The 6dF instrument
simultaneously observed up to 150 spectra over a 5.7\degrees\ field.
The science fibres of the instrument were 6.7 arcsec in diameter.  The
spectrograph is bench mounted in an enclosure inside the telescope
dome, which gives it increased instrumental stability compared to a
spectrographs mounted on a telescope (e.g. 2dF and the SDSS
spectrograph).  This increased instrumental stability maintains
consistent instrumental resolution, which is important when working
with Lick indices. The survey obtained spectra in the visual (V) and
red (R) wavelength ranges providing coverage from $\sim$4000 to
$\sim$8400\AA.  The spectra have 5-6\AA\ FWHM resolution in V and
9-12\AA\ resolution in R (see Jones et al. 2004 for more details).

The main science aims of the survey included: to measure the
luminosity function of near-infrared selected galaxies as a function
of environment and galaxy type (Jones et al. 2006), to measure the
clustering of galaxies, to produce detailed maps of the density and
peculiar velocity fields of the nearby universe (Jones et al. 2004;
Erdogdu et al. 2006) and to study the properties of the underlying
stellar population of the galaxies such as ages and chemical
abundances. It is this latter issue to which this paper is addressed.

The 6dFGS is particularly suited to the aims of the current study
because the selection criteria using 2MASS near-infrared photometry
provides a sample less biased toward bright, young galaxies than
surveys based on optical catalogues. The K band photometry also has
the useful property of being relatively insensitive to both metallicity 
the presence of dust (unlike optical bands), and is therefore an good 
tracer of mass. It also provides a useful tool in breaking the
age--metallicity degeneracy.  Finally, 6dFGS spectra are obtained from 
region 2 to 3 times larger on the sky than those used in the 2dFGRS and SDSS
surveys.  As a result, at the mean redshifts of the surveys, 
2dF, SDSS and 6dFGS fibres correspond to 2.6, 3.9 and 4.6 kpc/h 
respectively. The 6dFGS spectra therefore sample a 
significantly larger fraction of the total galaxy light than either 
2dF or SDSS spectra.


\subsection{Spectroscopic data}
\label{specsample}
The spectroscopic data presented here are from the 6dFGS First Data Release 
(Jones et al. 2004). Lick index, velocity dispersion and emission 
line-strength measurements 
were made on all galaxies with redshift quality~$\geq 3$ (the `reliable' 
redshifts; Jones et al. 2004). This gave a 
total of 39649 galaxies with 3859 of these observed more than once (420 
galaxies observed more than twice).  If a galaxy was observed more than once, 
only the highest signal-to-noise-ratio spectrum was used for the measurements. 
However, repeat observations were used to establish the errors (see Section 
\ref{error_measurement}). Only galaxies with median signal-to-noise ratios
$\ge$12 \AA$^{-1}$ were deemed suitable for age and metallicity 
determinations. Such a value lies at the lower limit of  signal-to-noise-ratio 
that can be used to derive ages and metallicities with precision.
The above selection criteria resulted in $\sim$7000 galaxies suitable for 
Lick index measurements. It should be noted that, as no morphological 
selection was made, the results presented here include galaxies of the 
full range of early and late types.\\

\subsection{Photometric data}
\label{samplephot}

\begin{figure}
\includegraphics[width=8cm,angle=0]{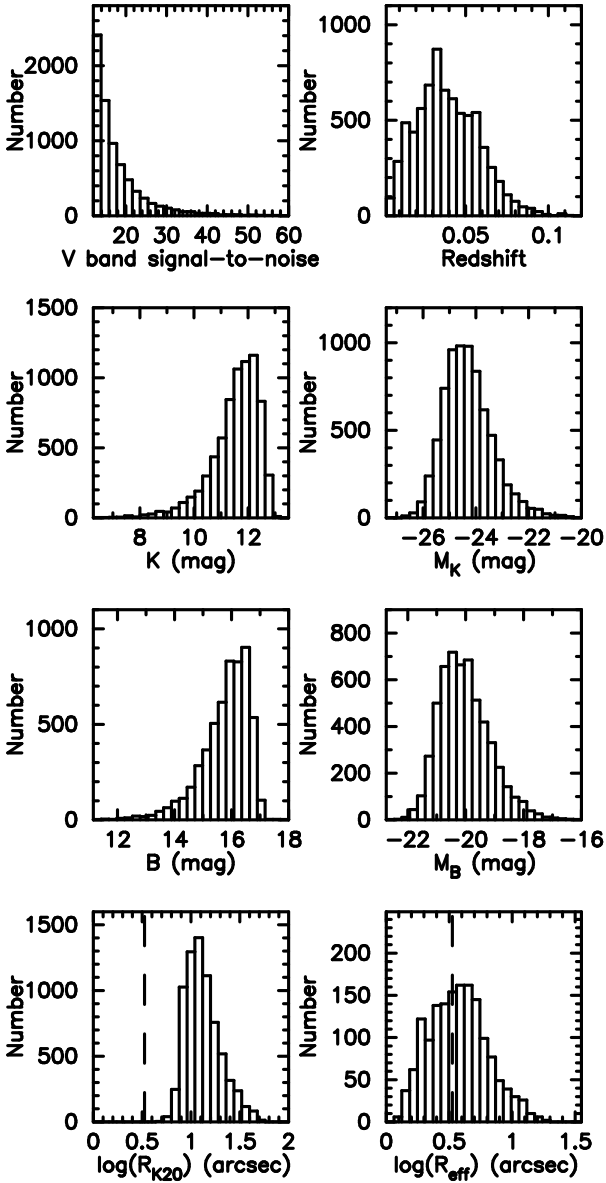}
\caption{The distributions of key parameters in the final sample of
$\sim$7000 galaxies presented here. The signal-to-noise-ratio per \AA\
is estimated as the median of the 6dFGS spectra. Apparent magnitudes
are from 2MASS and SuperCOSMOS data. Absolute magnitudes were
calculated using redshifts from the 6dFGS spectra and a Hubble
constant of 70 km s$^{-1}$Mpc$^{-1}$. K band 20th magnitude isophotal
radius is from 2MASS. Effective radii are also presented for a
($\sim$25\%) subsample of our 6dFGS galaxies. Dashed lines in the radius
plots represent the radius of the 6dF aperture (3.35 arcsec). }
\label{sample}
\end{figure}

The K band magnitudes used in this work were taken to be the K$_{tot}$
values estimated from the 2MASS data by Jones et al. (2004). We also
use the re-calibrated SuperCOSMOS B and R band magnitudes (Hambly et
al. 2001). Galaxy radii were taken as the 2MASS 20th magnitude per
arcsec$^2$ isophotal radius (R$_{K20}$; Jarrett et al. 2001a).  We also
use K band half-light (effective) radii (R$_{eff}$) for a
spectroscopically selected early-type galaxy subsample, as supplied by
Lachlan Campbell (private communication). Absolute magnitudes and
physical galaxy radii were calculated using distances based on the
measured redshifts and a Hubble constant of 70~km s$^{-1}$~Mpc$^{-1}$
(our median redshift is 0.035).  Errors on the distance estimates were
assumed to be a combination of recession velocity uncertainty and
peculiar motions. In order to be conservative, we assumed the combined
uncertainty to be $\pm$1000~km s$^{-1}$. The distributions of some key
parameters in the final sample are shown in Fig. \ref{sample}. The
sample covers of order 5 magnitudes in luminosity
($-26\le$M$_{K}$$\le-21$ mag and $-22\le$M$_B$$\le-17$ mag). This
corresponds to a dynamical mass range of more than two orders of
magnitude (from $<$10$^{10}$ to $>$10$^{12}$ M$_{\odot}$; Section
\ref{photo}). From the sample of $\sim$7000 galaxies suitable for
age--metallicity determination (Section \ref{specsample}) $\sim$6000
also had a full set of 2MASS and SuperCOSMOS photometry.

\section{Spectroscopic analysis}
\label{spectroscopicanal}

\subsection{Measuring emission lines}
\label{meas_em_lines}
We measured emission line-strengths for H$\alpha$, H$\beta$,
[NII]$\lambda$6548, [NII]$\lambda$6584, [OIII]$\lambda$4959 and
[OIII]$\lambda$5007 from the 6dFGS spectra. In order to measure the
\Halpha\ and [NII] emission, a de-blending option in the IRAF task
{\it splot} was used. This option allows the fitting of Gaussians to a
series of spectral lines and the measurement of the value of the
equivalent widths of each. The strengths of \Halpha\ and the two [NII]
doublet lines either side of it were measured in this way from
continuum divided spectra.  This process can lead to false
measurements where noise spikes or the continuum level are fit with
Gaussian profiles rather than real features.  By consideration of the
position and FWHM of the Gaussian fit to the feature, one can
reasonably discriminate between real emission features and false
detections.  If the line centre was more than 7~\AA\ away from the
features expected wavelength (given the 6dFGS redshift), the
measurement was discarded.  If the FWHM was well below the
instrumental resolution of the 6dF instrument ($<$ 5~\AA ) the
measurement was also discarded.  Finally, extremely large FWHM
($>$~12~\AA ) were discarded.  These large FWHM are mostly Gaussian
fits to features in the continuum, but may also have occasionally
included real features of broad-lined active galactic nuclei (AGN).

The equivalent widths of the two [OIII] emission lines at 4959 and 5007~\AA\ 
were measured using the procedure and band definitions of Gonzalez (1993). 
H$\beta$ emission, on the other hand, was estimated for each galaxy as the 
difference between the \emph{observed} 6dFGS H$\beta$ index value and the 
value predicted by our best-fit model solutions (see Section \ref{fitting}).

Within this work the sample is sometimes subdivided into `passive' (no
emission) and `emission' galaxies. Passive galaxies are defined as
those in which no emission lines are detected within the 6dFGS
wavelength range with greater than approximately 1-$\sigma$
significance. Emission line galaxies are therefore defined as those in
which at least one emission line is detected at greater than one sigma
confidence.  For H$\alpha$ this criterion corresponds to emission-line
equivalent widths of $\sim$1.5 \AA. Using this definition the sample 
contains 4640 passive galaxies and 2270 emission galaxies.
Unfortunately, a full analysis of morphological types is not available for 
the sample. However, visual inspection of a sub-sample of passive galaxies 
suggests that $\sim$80\% are early types, while amongst the emission galaxies 
a similar fraction are late types.

\subsection{Measuring velocity dispersion}
\label{measvd}

The method used to measure Lick indices from appropriately broadened
spectra and to correct the indices for the broadening effect of a
galaxy's velocity dispersion was based on that used by Stephen Moore (2001)

Measurements of the velocity dispersion in the 6dFGS galaxy spectra 
 were made using the IRAF task {\it fxcor} which carries out 
a Fourier cross--correlation between the object spectrum and a zero-redshift, 
zero-velocity dispersion template spectrum.  
The task outputs the radial velocity of the object 
spectrum and the FWHM of the Fourier correlation peak between the object and 
template spectra, which can be converted into a velocity dispersion.  
The template spectra used were three velocity standard stars observed with the
 6dF spectrograph in 2003 March by Craig Harrison (HR2574, HR3145 and 
HR5888, which are late G and early K giants).

The raw values of the velocity dispersion measured by {\it fxcor} need to be 
calibrated. To do this each of the standard star spectra were artificially 
broadened to a specific velocity dispersion using the IRAF task 
{\it gauss}. These artificially broadened spectra were then used to calibrate 
their (known) velocity dispersions against the value measured by {\it fxcor}. 
Towards lower velocity dispersions, as one approaches the instrumental velocity resolution ($\lesssim$100~km~s$^{-1}$), this 
calibration becomes less accurate. However, galaxies with such
low velocity dispersions require only very small corrections to their measured
Lick index values. Consequently, reliable index determinations can still be 
obtained in low velocity dispersion galaxies. More importantly, this method of 
measuring velocity dispersion fails on galaxies that are not well fit by the 
standard star templates. This mainly occurs in low signal-to-noise ratio 
galaxy spectra with young ages and/or strong emission lines. 

\begin{figure}
\center{\includegraphics[width=6cm,angle=-90]{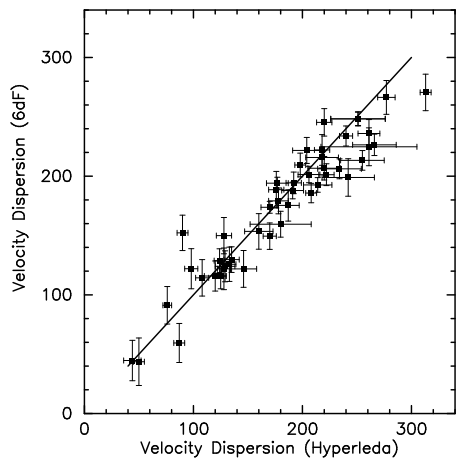}}
\caption{A comparison of velocity dispersion derived 
from 6dFGS spectra (in ~km s$^{-1}$)
with values from the literature (Hyperleda). Despite the large 6dF aperture we 
find an offset from the one-to-one line (shown as a solid line) of only 6 km s$^{-1}$ with an RMS 
scatter of 20~km s$^{-1}$. Such differences have only small effects on Lick 
index determinations.}
\label{vdisp_comp}
\end{figure}

A comparison of velocity dispersion values derived from the 6dFGS
spectra to values in the literature (Hyperleda) is shown in
Fig. \ref{vdisp_comp}. The 6dFGS values show good agreement with the literature values with an average offset of --6.1~kms$^{-1}$, and an RMS scatter of
20~kms$^{-1}$. Both deviation and scatter are greatest at high 
velocity dispersion, where we find the Hyperlead values to be 
consistently greater than the 6dFGS values. This is in agreement 
with Bernardi (2007) who found Hyperleda values greater than values 
derived from the SDSS at high velocity dispersion.

Given the large 6dF aperture (6.7 arcsec compared to a more typical 2
or 3 arcsec in Hyperleda) it is possible that these deviations are the
results of aperture effects. J{\o}rgensen et al. (1995) investigated
the effects of varying the aperture size in observations of velocity
dispersions in the inner regions (i.e.  R $\lesssim$ R$_{eff/2}$) of
early-type galaxies. Their results showed a weak power law dependence
of measured velocity dispersions with the size of aperture. We
therefore experimented with correcting velocity dispersions according
to J{\o}rgensen et al. (1995) in the sub-sample of 6dFGS galaxies for
which Hyperlead values were also available. As this subsample of the
data is heavily biased to nearby, early-type galaxies, it is 
well matched to the galaxies used to calibrate the J{\o}rgensen et
al. (1995) relation.  The results showed that while the offset was
reduced to +0.75~kms$^{-1}$, the scatter was increased to
23~kms$^{-1}$.  We also note that the trend for Hyperleda values to be
greater than values derived from our data remains unaffected by these
corrections.

As noted above, the J{\o}rgensen et al. (1995) correction is
calibrated for the inner regions of early-type galaxies.  However, for
our 6dFGS sample as a whole, the \emph{average} aperture is
$\sim$1~R$_{eff}$ (see Fig. \ref{sample}). The sample also contains
both early- and late-type galaxies. This suggests that the corrections
presented by J{\o}rgensen et al. (1995) are inappropriate for the 6dF
sample as a whole. For such a sample, it is instead informative to
consider the study of Gebhardt et al. (2000).  This analysis of the
variation in observed velocity dispersions (including the effects of
rotation) measured through circular apertures of varying radius in a
sample of 26 galaxies of mixed Hubble types shows that, within
5~R$_{eff}$, the sample exhibits little or no systematic variation of
the measured velocity dispersion with aperture size. 

As a result of the considerations above, we have elected to make no
corrections for aperture size to the measured velocity dispersion 
values within this work.\\

\subsection{Measuring Lick indices}
To make Lick index measurements suitable for direct comparison with 
single stellar population (SSP) models, it is 
necessary to broaden the spectra to the instrumental resolution of the 
Lick/IDS system.  The Lick/IDS FWHM 
instrumental resolution as a function of wavelength is described in Worthey \& 
Ottaviani (1997) where it is shown to vary from $\sim$8.5~\AA\ at the central
wavelength ($\sim$5000~\AA) to $\sim$10~\AA\ at the spectrum ends 
($\sim$4000 and 6000~\AA).

The calculation of the necessary broadening requires the measurement of the 
6dF instrumental resolution. This was measured from 6dF arc line spectra
obtained from the observational time-span of the 6dFGS First Data Release. 
The instrumental 
resolution of the 6dF spectra varies not only with wavelength but also with 
optical fibre number, so  measurements were made of arc lines at various 
wavelengths, \emph{for all fibres individually}.  A bivariate polynomial was 
fit to the measured FWHM spectral resolution as a function of both wavelength 
and fibre number.  The amount of broadening required to match the Lick/IDS 
resolution could then be estimated and each 6dF spectrum was broadened 
to this resolution.

\subsection{Velocity dispersion correction}
The index values given in stellar population models do not include the effect 
of the velocity dispersion broadening present in galaxy spectra.  It is 
therefore necessary to introduce a 
correction to the observed galaxy index values to account for this effect. 

To measure the velocity dispersion correction required for each Lick
index, we measured the indices of a range of artificially broadened
stellar spectra.  These values of the Lick indices measured in the
broadened spectra were then compared to those measured in the
un-broadened spectra to determine the index corrections for a range of
velocity dispersions.

For each index measured in Angstroms (e.g. the
Fe and Ca indices) a multiplicative correction was defined by:

\begin{equation}
	\rm Index\ correction = \frac{Index_{\sigma=0}}{Index_{\sigma}}
\end{equation}

\noindent While, for each index measured in magnitudes (e.g. CN$_1$,
CN$_2$, Mg$_1$, Mg$_2$), an additive corrections was defined
by:

\begin{equation}
	\rm Index\ correction = Index_{\sigma=0}-Index_{\sigma}
\end{equation}

\noindent Where index$_{\sigma = 0}$ is the value of the index in the 
raw stellar spectrum (i.e. with no broadening) and index$_{\sigma}$ 
is the index value after broadening of the stellar spectrum to a 
specific velocity dispersion ${\sigma}$.

For each index, quadratic lines were fitted to the velocity dispersion versus 
index correction ratio data.  
The index correction required for a particular galaxy could then be 
determined from the measured velocity dispersion and the quadratic function. 

\subsection{Estimating errors}
\label{error_measurement}
Estimating errors for spectroscopic measurements is notoriously difficult,
mainly because there are several sources of error that can not be well 
quantified (e.g. sky-subtraction, poor flux calibration etc). Therefore, 
instead of trying to propagate variance arrays, comparisons of the values 
from repeated 6dFGS observations of the same galaxy were used to quantify 
this error.  This method was used for calculating the error in Lick indices 
and velocity dispersions, as well as in \Halpha\ and [NII] emission 
equivalent-widths.

\begin{figure}
\centerline{
\includegraphics[width=4.2cm,angle=0]{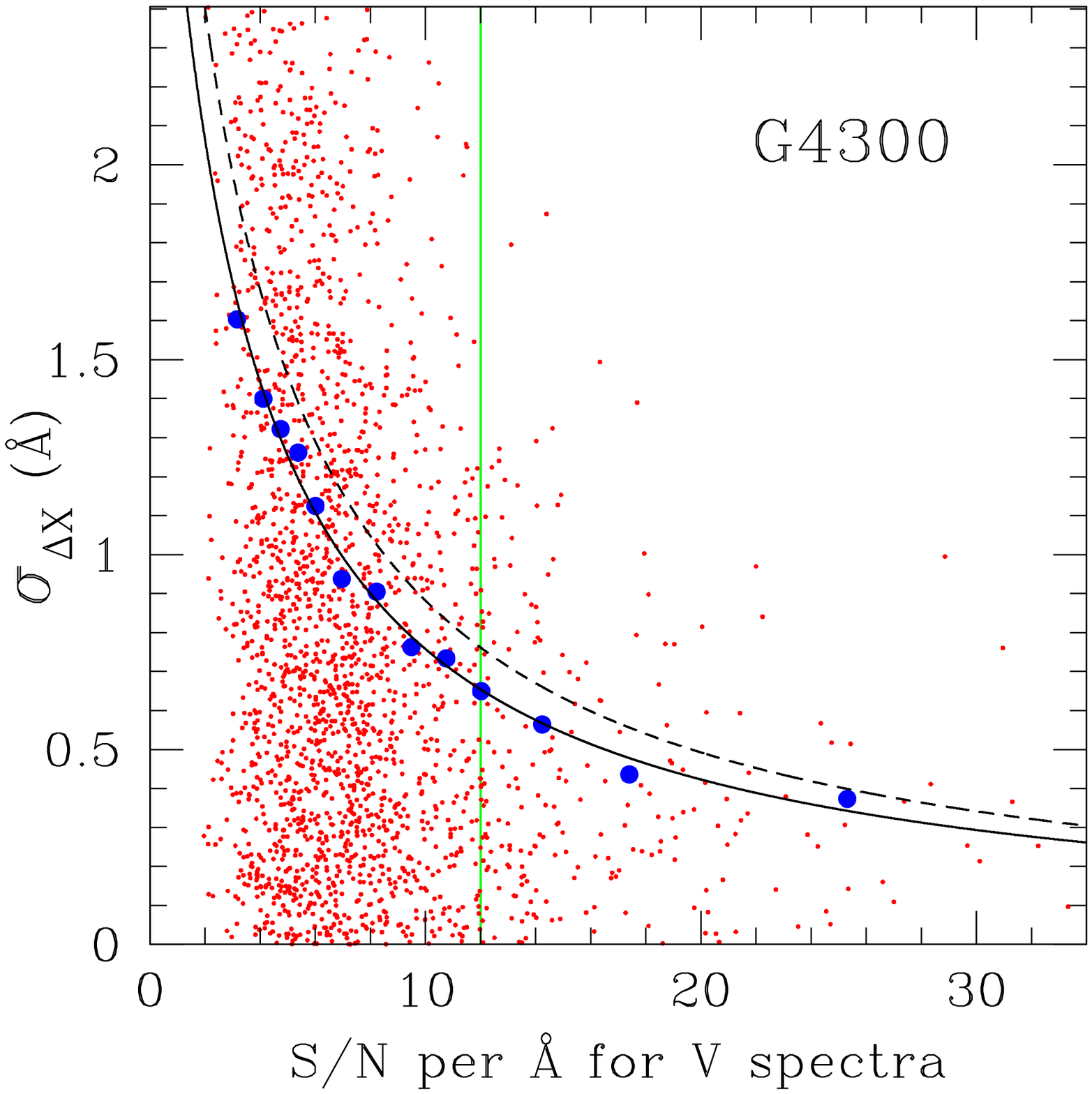}
\includegraphics[width=4.2cm,angle=0]{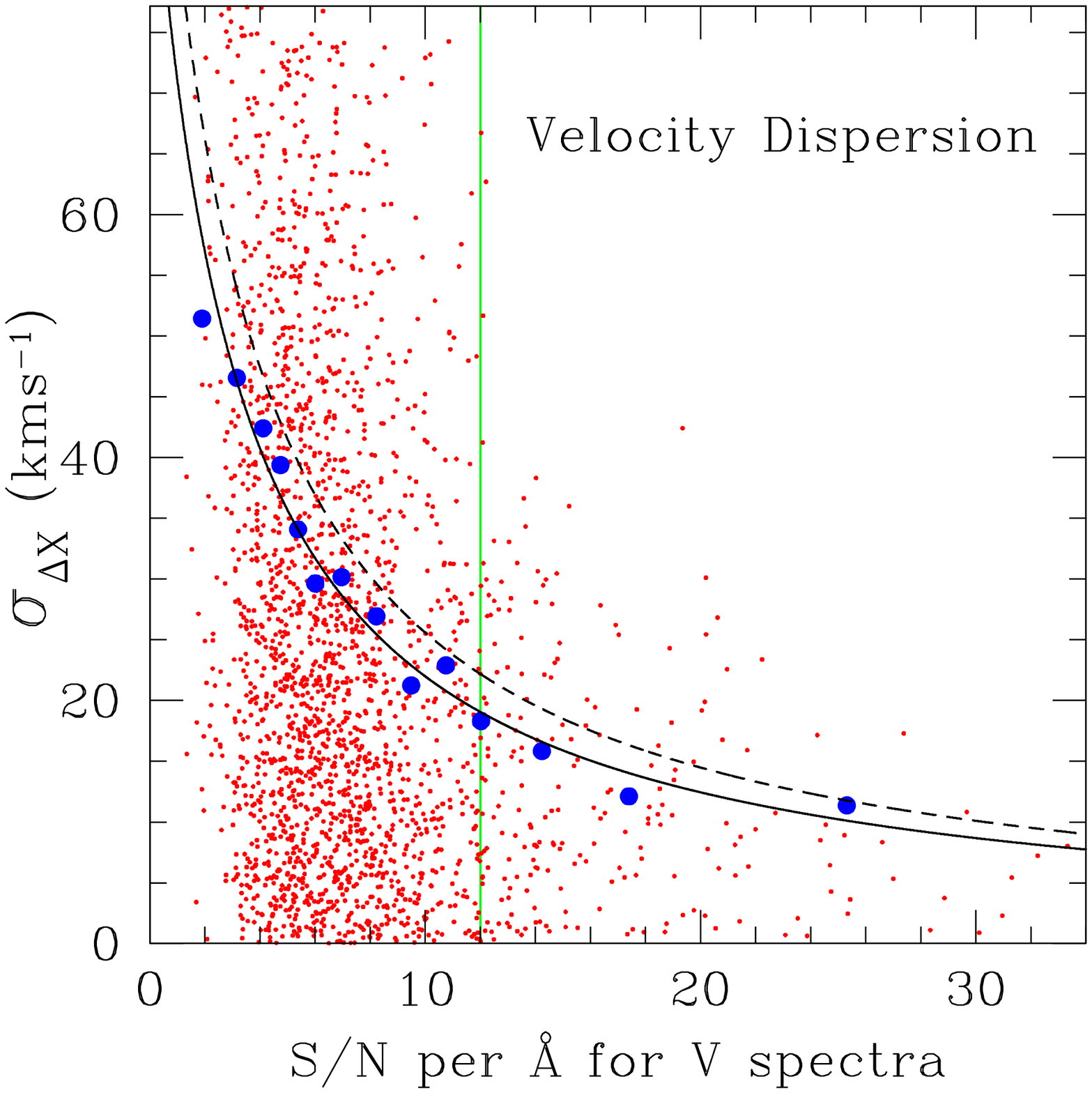}
}
\centerline{
\includegraphics[width=4.2cm,angle=0]{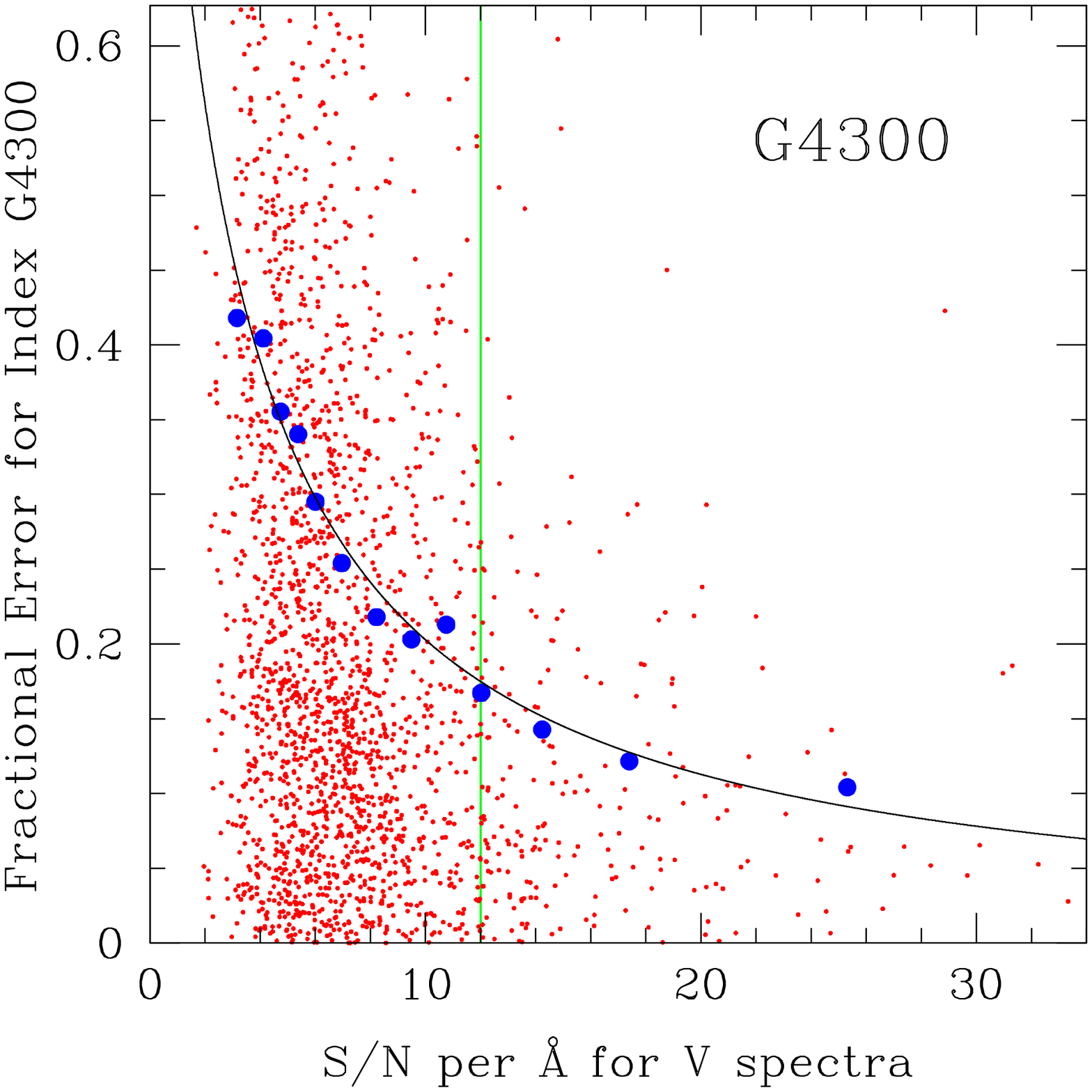}
\includegraphics[width=4.2cm,angle=0]{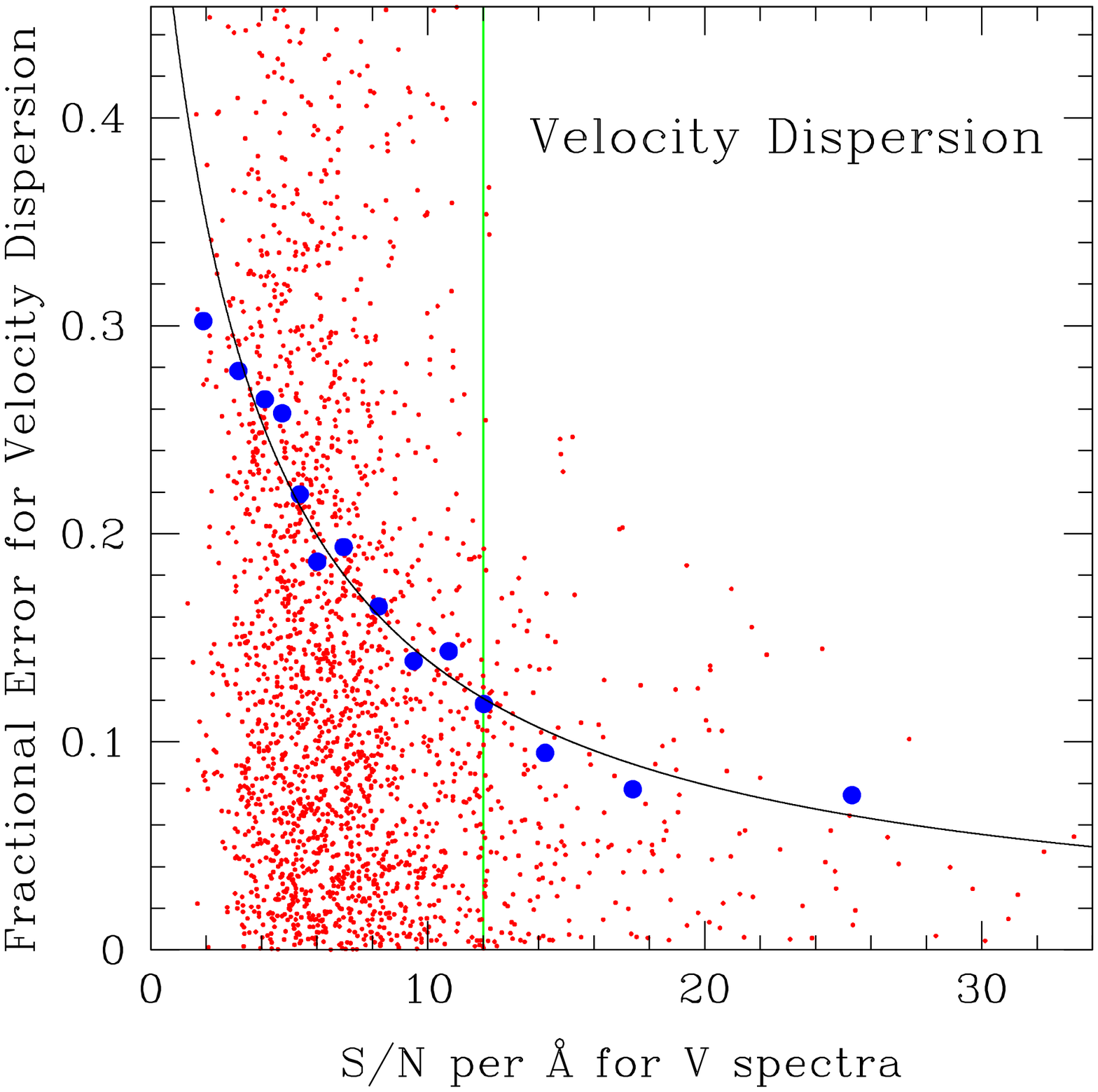}
}
\caption{{\bf Top}:The measured relationship between RMS error and
signal-to-noise ratio from repeated observations for the G4300 index
(left hand plot) and for velocity dispersion (right plot). The small
points are the individual differences between repeated observations
for each galaxy. The large points are the RMS binned values of these
points after $2\sigma$ clipping.  The solid line is the least squares
fit to the bin values.  The dashed line is the relationship between
error and signal-to-noise ratio after correcting for the clipping used
in forming the RMS bins (see Section \ref{error_measurement}) The
vertical line marks the lowest signal-to-noise ratio (12
\AA$^{-1}$)used in the present study. {\bf Bottom}: Same as top plots except 
the {\it fractional} error is shown.}
\label{error_SNR}
\end{figure}

In the First Data Release of the 6dFGS, 3859 galaxies were observed
more than once. For all repeat observations the differences in values
were grouped into bins based on the median signal-to-noise ratio of
the galaxy spectra.  The signal-to-noise ratio of the lower of the two
observations was used for this binning, as this observation would be
the cause of most of the error.  If a galaxy had been observed more
than twice then the two highest signal-to-noise ratio spectra were
used.  The RMS of each signal-to-noise ratio bin was determined and
used as the $\sigma_{\Delta x}$ value.  To remove the biasing effect
of large outliers in the bins only the lowest 95.45\% of the values
(the $2\sigma$ limit) in each bin was used to calculated the RMS
value.  This requires that the RMS be scaled up by 1.164 to take into
account this clipping.

A function was fitted to the RMS of the bins against the signal-to-noise ratio.
The fitting function used was a  modified rectangular hyperbola of the form 
$y = a/(x+b)$ (`a' and `b' the fitted coefficients). The error in any measured 
quantity for a galaxy could then be estimated using the signal-to-noise ratio 
of the spectrum and the corresponding error from the fitted function. Examples 
of this procedure can be seen in Figure \ref{error_SNR}. We note that the 
majority of the sample with repeat measurements were re-sampled due to the 
low signal-to-noise of the original observation.

\subsection{Fitting indices to models}
\label{fitting}

\begin{figure}
\includegraphics[width=8cm,angle=-90]{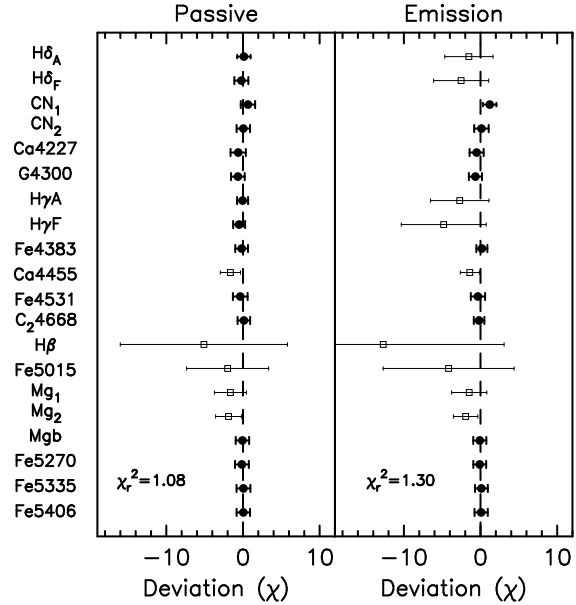}
\caption{Mean residuals of indices to the final best-fit SSP models are 
given in units of observational error (i.e. $\chi$). Error bars represent 
the RMS scatter about the mean. Solid symbols represent indices included in 
the fits, while open symbols represent those indices excluded from the fits.
The results for fits to both passive (left) and emission galaxies (right) are 
shown.
The average {\it reduced}--$\chi^2$ of indices included in each subsample 
($\chi^2_r$) is 
indicated in each plot.}
\label{chis}
\end{figure}

We used the $\chi^2$-fitting procedure of Proctor \& Sansom (2002)
(see also Proctor et al. 2004a,b and Proctor et al. 2005) to measure the 
derived parameters: log(age), [Fe/H], [Z/H] and [E/Fe] (a proxy for 
the `$\alpha$'--abundance ratio; see Thomas et al. 2003 for details).
Briefly, the technique for deriving these parameters
involves the simultaneous comparison of as many observed indices as possible 
to models of single stellar populations (SSPs). The best fit is found by 
minimising the square of the deviations between 
observations and models in terms of the observational errors (i.e. $\chi^2$).
The rationale behind this approach is that, while all
indices show some degeneracy with respect to each of the derived parameters, 
each index does contain \emph{some} information regarding
each parameter. In addition, such an approach should be relatively robust
with respect to many problems that are commonly experienced in the
measurement of spectral indices and their errors. These include poor flux 
calibration, poor sky subtraction, poorly constrained velocity dispersions, 
poor calibration to the Lick system and emission-line contamination. This 
robustness is of particular importance in the analysis of large numbers of 
pipe-line reduced spectra such as those of the 6dFGS which cannot be flux 
calibrated and so are not fully calibrated to the Lick system. The method 
is also relatively robust with respect to the uncertainties in the SSP models 
used in 
the interpretation of the measured indices; e.g. the second parameter effect 
in horizontal branch morphologies and the uncertainties associated with the 
Asymptotic Giant Branch. It was shown in Proctor et al. 
(2004a) and Proctor et al. (2005) that the results derived using the $\chi^2$ 
technique are, indeed, significantly more reliable than those based on only a 
few indices.

In order to carry out the comparison of observations to SSP models it was first
necessary to select and interpolate the models provided in the literature.
Interpolations are required because models in the literature are presented 
at only five or six discrete metallicities over the $\sim$2.5~dex range. 
We experimented with a number of different models (Bruzual \& 
Charlot 2003 (hereafter BC03); Thomas, Maraston \& 
Bender 2003; Thomas, Maraston \& Korn 2004 and Korn, Maraston \& Thomas 
2005 (hereafter KMT05). The results presented 
here are those from the models of KMT05, as these provide 
the necessary coverage in age (0.1 to 15 Gyr) and metallicity ([Z/H] from 
--2.3 to +0.67),
as well as including a detailed modelling of the effects on indices of 
varying `$\alpha$'--abundance ratios. However, fits were also obtained to the 
other model sets for comparison.

The process by which the best fits were obtained was iterative. First, fits 
were obtained for all galaxies using {\it all} the available indices. The 
pattern of deviations from the fit so obtained was then used to identify 
indices that matched the models poorly  (see Fig. \ref{chis}). Due to the 
problem of emission-line filling of the H$\delta$, H$\gamma$ and H$\beta$ 
Balmer lines in emission galaxies, we considered `passive' and 
`emission' galaxies (Section \ref{meas_em_lines}) separately. 
In Fig. \ref{chis} emission galaxies show the expected large deviation between 
observed and best fit model values of Balmer indices and Fe5015 (which is 
also emission-line affected). These indices were therefore omitted from the 
fitting procedure in emission galaxies.

Perhaps surprisingly, H$\beta$ and Fe5015 also deviate significantly in the 
`passive' galaxies. We take this to indicate that, despite the rather severe 
definition of emission galaxies outlined in Section 
\ref{meas_em_lines}, at least some of the passive galaxies contain low-level 
emission that has escaped detection during our emission line measurements. 
Indeed, this was the main motivation for the stringent definition of 
emission-galaxies.

To minimise the effects of this low-level emission on our age and metallicity 
estimates, age sensitive H$\beta$ and metallicity sensitive 
Fe5015 indices were also excluded from the fits to passive galaxy data.

In both passive and emission galaxies Ca4455 was found to fall below 
best-fit model values. This effect has been noted in previous studies 
(e.g. Proctor \& Sansom 2002; Clemens et al 2006). The flux-calibration 
sensitive Mg$_1$ and Mg$_2$ indices were also 
found to deviate from best fit model values. This was an anticipated effect, 
as 6dFGS spectra are only approximately flux calibrated and Mg$_1$ and Mg$_2$ have widely space pseudo-continuum bands. 
Ca4455,  Mg$_1$ and Mg$_2$ were therefore excluded from the fitting 
procedure. Fortunately, most of the information lost by the exclusion of the 
Mg$_1$ and Mg$_2$ indices is captured in the Mgb index that is not sensitive 
to flux-calibration issues.

\begin{figure}
\includegraphics[width=7cm,angle=-90]{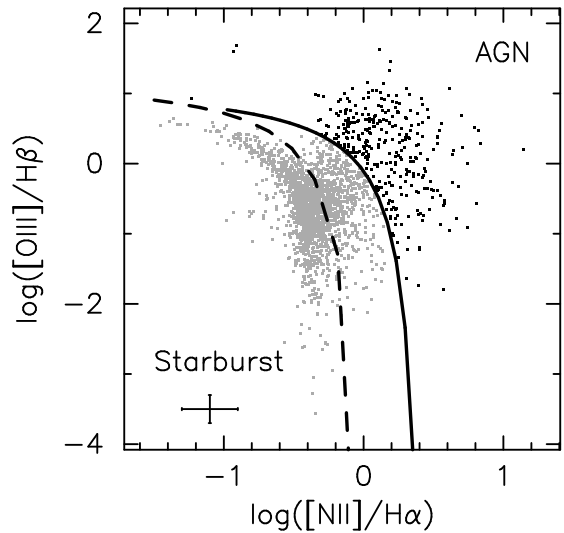}
\caption{The [OIII]/H$\beta$ vs [NII]/H$\alpha$ emission line diagnostic 
diagram. NII, H$\alpha$ and [OIII] were measured directly (see Section 
\ref{meas_em_lines}). H$\beta$ emission was estimated as the difference between
observed and  best-fit SSP model values of the H$\beta$ index (see Section 
\ref{fitting}). The average errors in the line-ratios are shown in the 
bottom-left of the plot. The solid line is from Kewley et al. (2001) and 
represents the theoretical dividing line between star-burst galaxies 
(grey points) and AGN (black points). 
The dashed line is the equivalent line from Kauffmann (2003a). Conservatively, 
we use the Kewley line to identify AGN.}
\label{emission}
\end{figure}

Once poorly fitting indices had been identified and omitted, the fitting of the
remaining indices was carried out using a clipping procedure in which indices 
deviating from the model fit by more than 3$\sigma$ were excluded, and the 
fitting procedure performed again. This results in an average of less than one 
additional index per galaxy being clipped from the fits. Most of the clipped 
indices could be associated with contamination from features such as the 
5577~\AA\ sky-line. The 
resultant fits are based on between 10 and 15 indices and show good agreement 
between observed and best-fit values (Fig. \ref{chis}). Indeed, the 
reduced-$\chi^2$ values given in Fig. \ref{chis} ($\chi^2_r$) were 1.08 and 
1.30 for passive- and emission-galaxies respectively. This 
clearly indicates that our data are well represented by the models and that 
our errors reflect well the differences between 
observations and models.

\subsection{Errors in derived parameters}
\label{errdps}
For each galaxy in the sample, errors in log(age) and [Z/H] were estimated 
using 50 Monte-Carlo realisations of the best-fit model indices, assuming the 
observational errors for the galaxy in question. We again note that the 
reduced-$\chi^2$ values of both passive- and emission-galaxies were close to 
one, indicating that the assumed errors reflect the average 
differences between observations and models quite well. The RMS scatter in the 
log(age) and [Z/H] values of the 50 realisations was therefore taken as the 
individual galaxy error. However, in an effort to better understand the 
detailed behaviour of the errors, deviations from the best-fit log(age) and 
[Z/H] values of all 50 realisations of each galaxy were retained and are 
presented in Section \ref{AZERR}. 
For the purposes of the following analysis it is important that the derived 
ages are well defined. Therefore, $\sim$~1000 low signal-to-noise galaxies 
whose error in log(age) exceeded 0.3~dex were deemed unreliable and were 
excluded from the analysis. The average errors for the remaining 
galaxies were $\sim$0.15~dex in both log(age) and [Z/H].

\section{Results}
\label{results}
\subsection{Emission line galaxies}
\label{emlingal}
Emission lines can be used to separate star-burst and active galactic nuclei 
(AGN) galaxies using the [OIII]$\lambda$5007/H$\beta$--[NII]/H$\alpha$ 
diagnostic. Values of the [OIII]$\lambda$5007, [NII] and H$\alpha$ emission 
strengths were estimated from the 6dFGS spectra in a traditional manner 
(Section \ref{meas_em_lines}). The strength of H$\beta$ emission, however, 
was taken to be the deviation between the measured value of the H$\beta$ index 
and the best-fit SSP model value. Due to the relatively small observed 
dynamical range of H$\beta$ in absorption compared to that observed in 
emission, this is a fairly robust
estimate of H$\beta$ emission. This is evidenced by the classic `Y'-shape 
evident in the diagnostic plot of [OIII]$\lambda$5007/H$\beta$ against 
[NII]/H$\alpha$ shown in in Fig. \ref{emission}. The position of the line 
dividing  star-burst galaxies from AGN varies between studies in the literature.
In Fig. \ref{emission} we show the lines given by Kewley et al. (2001) (solid 
line) and Kauffmann (2003a) (dashed line). In order to ensure reliable 
identifications of AGN we use the definition of Kewley et al. in the following.
About 2200 of the 2700 emission galaxies in our sample have reliable 
estimates of all four of these emission lines, 300 of these lying in the 
region of the diagram associated with AGN as defined by Kewley et al. (2001).\\

\subsection{Age and metallicity}
\label{agemet}
In this section we outline the results of our age and metallicity 
determinations using Lick indices. 

\begin{figure*}
\includegraphics[width=10cm,angle=-90]{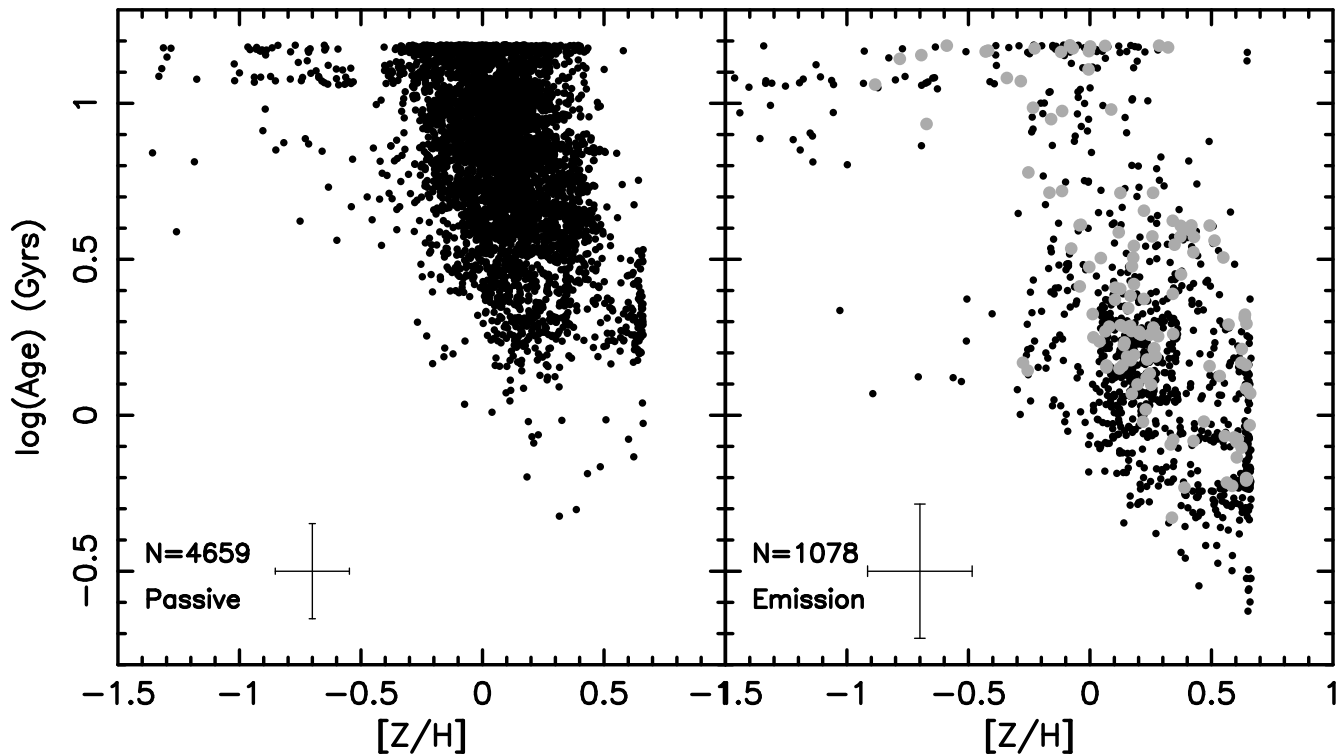}
\caption{The ages and metallicities derived from the 6dFGS spectra
separated by emission type (see Section \ref{meas_em_lines}). Grey
points in the right-hand panel represent the AGN identified in
Fig. \ref{emission}. Average errors are shown in each plot.  
The overall loci of
the points show that passive and emission galaxies largely follow
similar trends between age and metallicity -- albeit with emission
galaxies tending to significantly younger ages.}
\label{agez}
\end{figure*}

The distribution of ages and metallicities are presented in Fig. \ref{agez}. 
The sample has been sub-divided into passive- and emission-galaxies 
as described in Section \ref{meas_em_lines}. Emission-galaxies classified 
as AGN (Section \ref{emlingal}) are shown as grey symbols, the remainder are 
plotted in black.

Before interpreting this plot, we must note the apparent clusterings and 
`zones-of-avoidance', the clearest examples of which  are evident in the 
emission galaxy plot. These are the inevitable result of linear interpolations
in non-rectilinear spaces. 

Returning attention to the actual results of our age and metallicity 
determinations; similar trends of increasing [Z/H] with decreasing log(age) 
are evident in both of passive and emission galaxies  -- albeit with emission 
galaxies tending to younger ages. Galaxies classified as possessing AGN in 
Section \ref{emlingal} also often possess young central ages. 

\begin{figure}
\includegraphics[width=5cm,angle=-90]{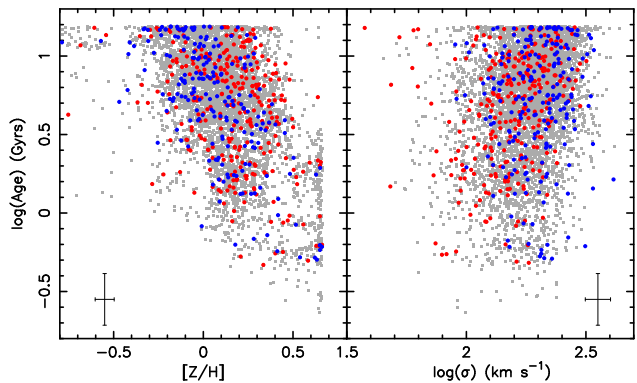}
\caption{Age--metallicity and age--log($\sigma$) are shown with colour
representing galaxy (apparent) size. Blue points represent the
smallest 5\% of our sample, red points the largest
5\%. Age--metallicity and age--log($\sigma$) are shown to be largely
independent of galaxy size.}
\label{aperture}
\end{figure}

We investigate the effect of aperture size on derived parameter 
using Fig. \ref{aperture}. In these plots, the galaxies
with the largest and smallest sizes (R$_{K20}>$37.5 arcsec and
R$_{K20}<$7.5 arcsec) are shown as red and blue points
respectively. The blue points therefore represent the galaxies in
which the 6.7 arcsec AAOmega aperture encompasses the largest fraction
of galaxy light (on average $\sim$3.0~R$_{eff}$), while the red points
represent galaxies in which the smallest fraction of galaxy light
($\sim$0.5~R$_{eff}$) is encompassed by the fibre.  The observed
age--metallicity and age--velocity dispersion relations can be seen to
be similar in these two extremes, with the most noticeable difference a
slightly higher [Z/H] (by $\sim$0.1~dex) in large galaxies. We 
therefore find aperture-size effects to be generally small.

Due to the large, circular aperture of the 6dFGS, a
direct comparison with the literature, which are mainly based on
long-slit observations of galaxy centres, is not straight
forward. Nevertheless, we note that the observed trend is at least
qualitatively in agreement with the trends observed in previous
studies (Trager et al. 2000; Proctor \& Sansom 2002; Mehlert et
al. 2003; Gallazzi et al. 2005; Collobert et al. 2006). The young
central ages of AGN are also consistent with studies in the literature
showing that star-bursts are often found to accompany AGN activity
(e.g. see Cid Fernandes et al. 2004 and references therein).

An intriguing feature of Fig. \ref{agez} is the presence of a few old galaxies 
with extremely low [Z/H] values, ($\lesssim$--0.5 dex). These galaxies, 
present in both passive and emission samples, possess a range 
of velocity dispersions. 
Visual inspection of both spectra and photometry generally reveals nothing 
exceptional about these galaxies. Further analysis of these galaxies are beyond the scope of this paper, but we note that their low numbers result in them 
having no impact on the results and conclusions of the present paper.

As in these literature studies, the slope in the age--metallicity relation 
exhibited by our data is similar to the slope of the age--metallicity 
degeneracy (Worthey 1994; `the 3/2 rule').
Before comparing these results with the photometry, it is therefore clearly 
important to establish that the age--metallicity degeneracy has been broken. 
To this end, in the following section we consider the errors in the 
derived parameters.

\subsection{Errors in age and metallicity}
\label{AZERR}
As described in Section \ref{errdps}, errors in age and metallicity
were characterised by considering the results of the 50 Monte-Carlo
realisations of each of the 7000 galaxies in our sample.  Fits using
both the index combinations shown in Fig. \ref{chis} (with and without
Balmer lines) were carried out for the purposes of this analysis (a
total of approximately one million realisations). The differences
between the input-model values of log(age) and [Z/H] and those of each
of the realisations were calculated. These were combined in a number
of bins depending upon the age--metallicity of the input models.

The analysis was also carried out using the \emph{observed} index values as 
the inputs to the Monte-Carlo realisations (rather than the best-fit model 
values). No significant quantitative or qualitative differences were found.

\begin{figure*}
\includegraphics[width=17cm,angle=0]{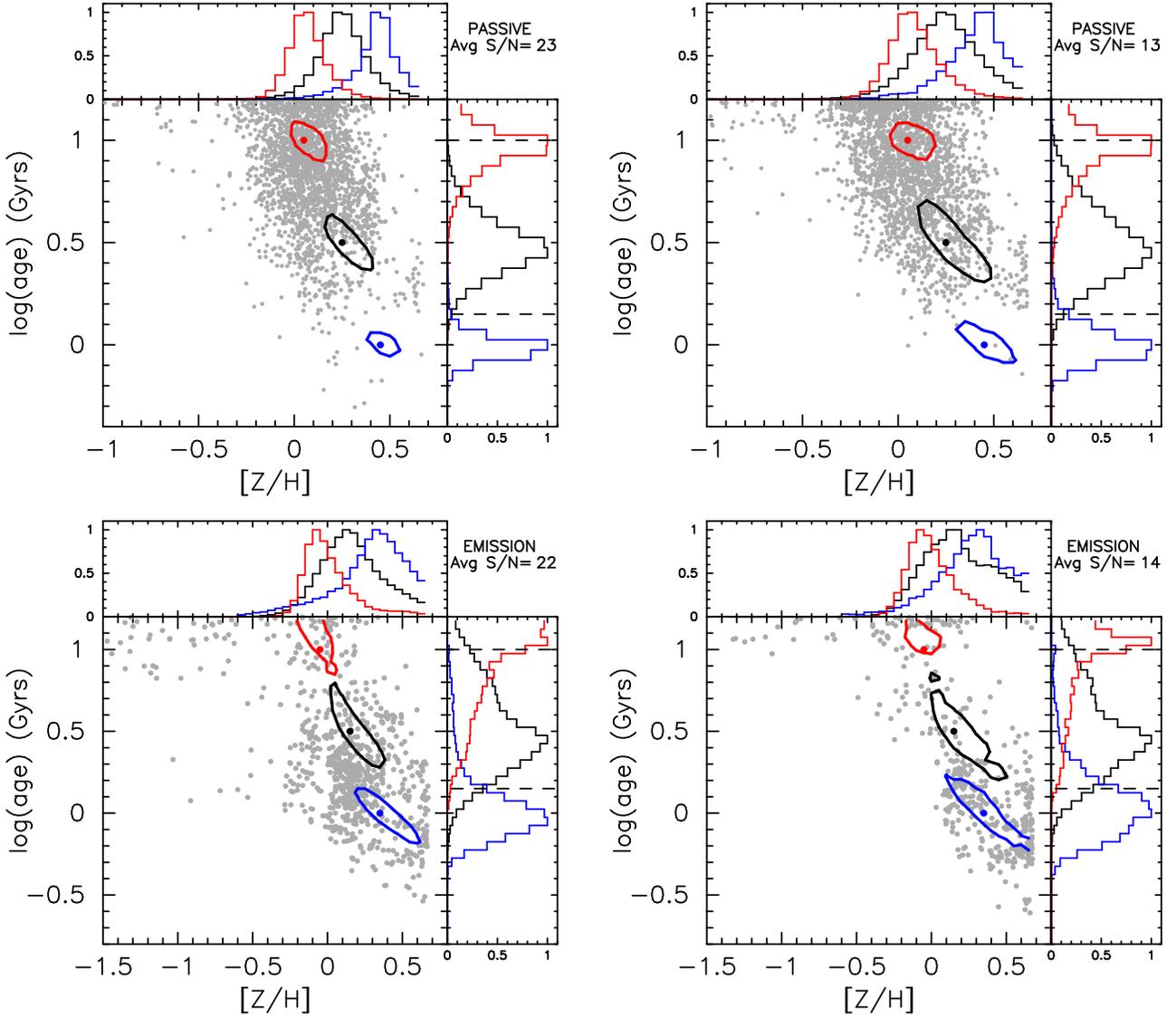}
\caption{Error contours derived from the Monte-Carlo realisations of
our galaxy sample are shown in three age--metallicity bins. Two
signal-to-noise ratio subsamples are shown; S/N$>$16~\AA$^{-1}$ (left)
S/N and $<$16~\AA$^{-1}$ (right).  Results are given for both passive-
(top) and emission-line (bottom) galaxies using the appropriate index
combinations (see Section \ref{fitting} and Fig. \ref{chis}). The data
for galaxies of appropriate emission characteristics and
signal-to-noise-ratio are shown as grey points in each plot. Dashed
lines in marginal distributions represent the upper and lower limits
used to define the `young' and `old' samples respectively.  Although the
age--metallicity degeneracy is still present, its effects are
minimised by the relatively small errors.}
\label{error}
\end{figure*}

Fig. \ref{error} shows the results of this analysis as 1-sigma
confidence contours for three age--metallicity bins. Each contour in
these plots is based on tens of thousands of individual realisations
of the best-fit galaxy data. The contour levels were defined as the
iso-densities corresponding to e$^{-1}$ of the peak values (as
expected for the 1-sigma contour of a two-dimensional Gaussian
distribution). The use of the iso-density peak in estimating the
extent of 1-sigma confidence contours results in a small uncertainty
in the estimates. However, we estimate these to be of order 2\% and 
therefore insignificant.

Results of the error analysis for both passive galaxies (in which 
Balmer indices were included in fits) and emission galaxies (in which Balmer 
indices were \emph{excluded} from fits) are shown separately. We recall that 
H$\beta$ was excluded from all the fits. Fig. \ref{error} also shows the data 
divided into two signal-to-noise ratio regimes (S/N$<$16~\AA$^{-1}$ and 
S/N$>$16~\AA$^{-1}$). Marginal distributions in log(age) and [Z/H] are 
shown at the edges of each main plot. We note that the positions of the 
contours in 
Fig. \ref{error} were chosen for clarity only. To get a full picture one must 
imagine the whole surface of the log(age)--[Z/H] plane populated by such 
contours.

Fig. \ref{error} shows that, as might be expected, error estimates
increase with decreasing signal-to-noise-ratio. For passive galaxies,
the analysis reveals log(age) and [Z/H] errors of $\sim$0.1~dex in
galaxies with S/N above $\sim$16.  This falls to $\sim$0.2~dex in
galaxies with S/N below $\sim$16. The errors of the emission-galaxies
(in which all Balmer lines are excluded from the fits) are somewhat
larger, but are generally still of order 0.2~dex.  It is therefore
clear that while the age--metallicity degeneracy is still present (as
evidenced by the sloped elliptical error contours in Fig.
\ref{error}), the magnitude of the errors is sufficiently small for
its effects to be negligible in our results - even in galaxies fit
without Balmer lines. This is the result, and main advantage, 
of using large numbers ($>$10) indices in the determination of ages and
metallicities. The good age resolution achieved is emphasised
in the marginal distributions of Fig. \ref{error}, in which galaxies
older than 10~Gyr (1.0~dex), and galaxies younger than 1.5~Gyr
(0.15~dex) (delimited by dashed lines in the marginal distributions)
are largely uncontaminated by galaxies with ages of 3~Gyr
(0.5~dex). Since the following analysis concentrates mainly on the
very oldest and very youngest galaxies, these provide confidence that
results do not suffer significantly from the effects of the
age--metallicity degeneracy.

The final sample for which velocity dispersions and reliable age/metallicity 
estimates were measured consists of 4500 passive galaxies and 1000 emission 
galaxies.\\

\subsection{Photometric properties and dynamical masses}
\label{photo}
The combination of 6dFGS, 2MASS and SuperCOSMOS data permits the comparison 
for spectroscopic age and metallicity determinations to the B, R and K band 
photometric data for
some 6000 galaxies. Specifically, by combining the photometry with the velocity
dispersion measures from 6dFGS spectroscopy, we are able to investigate
trends with age and metallicity in B, R and K band mass-to-light-ratios. 

This analysis is carried out using a \emph{dynamical} mass (M$_{dyn}$) 
calculated as:

\begin{equation}
  \rm M_{dyn} = \frac{C \sigma^2 R_{eff}}{G}
\label{masseq}
\end{equation}

\begin{figure}
\includegraphics[width=8cm,angle=-90]{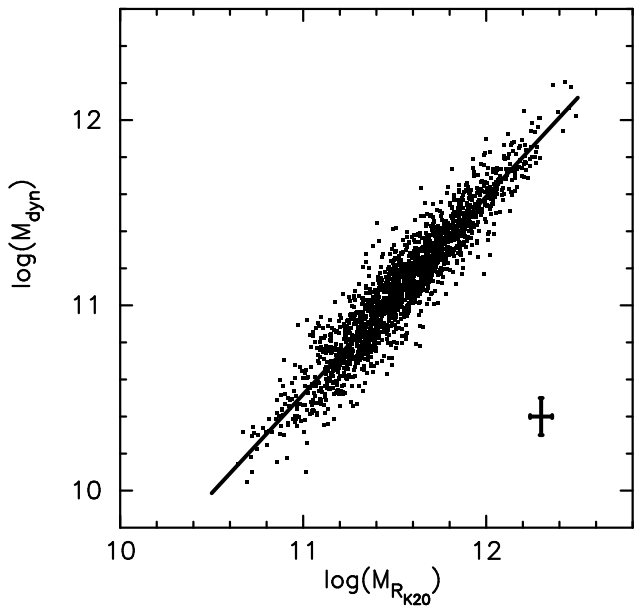}
\caption{A comparison of dynamical mass (M$_{dyn}$; Equation
\ref{masseq}) with M$_{\mathrm{R_{K20}}}$ (the mass estimate using
R$_{\mathrm{K20}}$ in place of R$_{\mathrm{eff}}$). The solid line shows
the relation M$_{\mathrm{dyn}}$=1.068M$_{\mathrm{RK20}}$--1.223. The rms
scatter about this line is 0.12~dex.}
\label{mass_comp}
\end{figure}

\noindent Where $\sigma$ is the central velocity dispersion,
R$_{\mathrm{eff}}$ is the half-light radius and the constant \rm{C} has
a value of 5.0 (Cappellari et al. 2006).  Unfortunately half-light
radii are only available for 25\% of our sample, while K band 20th
magnitude isophotal radii (R$_{\mathrm{K20}}$) are available for the entire
sample. A calibration was therefore carried out by using Equation
\ref{masseq} to calculate masses for the galaxies in the 25\%
sub-sample (which also possess much more accurate, aperture
corrected, velocity dispersion estimates), and comparing them with mass
estimates using R$_{K20}$ (instead of R$_{\mathrm{eff}}$) in the same
equation. A plot of the comparison is shown in
Fig. \ref{mass_comp}. The correlation has equation
M$_{\mathrm{dyn}}$=1.068M$_{\mathrm{R_{K20}}}$--1.223 with only 0.12~dex of
scatter. Given error estimates in M$_{dyn}$ and M$_{\mathrm{R_{K20}}}$ of 0.06
and 0.10~dex respectively, the correlation is clearly extremely
good. Masses quoted throughout the remainder of this work are
therefore those based on R$_{K20}$ corrected as detailed.


As well as the photometry of 2MASS and SuperCOSMOS, we also use the
photometric results from the SSP models of Bruzual \& Charlot
(2003). Some key values from the Bruzual \& Charlot models are shown
in Table \ref{photometry}. We use these models almost exclusively to estimate
differential properties; e.g. the rate of change of
mass-to-light-ratio with log(age) in populations of given [Z/H].  It
should be noted that the rates of change of the photometric properties
are reasonably constant with metallicity and age, as long as age is
expressed in logarithmic form. This clearly indicates that a
differential approach is a robust use of the model values.  We shall
flag the one occasion in which the models are used in a
non-differential manner.

\begin{table}
\begin{centering}
\begin{tabular}[b]{|c|c|c|c|c|c|}  
\hline
Age   &log(age) & [Z/H] &B&R&K\\
(Gyrs)& (dex)  & (dex) & (mag) & (mag) & (mag)\\
\hline
1     &0.000 &-0.4 & 5.27   &  4.40   & 2.67   \\
3     &0.477 &     & 6.61   &  5.33   &	3.24   \\
11.7  &1.105 &     & 7.92   &  6.46   & 4.15   \\
\hline		     	           		  		      
1     &0.000 & 0.0 & 5.53   &  4.52   & 2.61   \\
3     &0.477 &     & 6.94   &  5.55   &	3.20   \\
11.7  &1.105 &     & 8.32   &  6.73   & 4.15   \\
\hline		     	           		  		      
1     &0.000 & 0.4 & 5.91   &  4.72   & 2.27   \\
3     &0.477 &     & 7.29   &  5.76   &	3.03   \\
11.7  &1.105 &     & 8.79   &  7.01   & 4.04   \\
\hline
\end{tabular}
\caption{ Examples of B, R and K magnitudes from BC03 for single
stellar population models of varying age and metallicity are
presented.  Note that these values are used almost exclusively in a
differential manner (see Section \ref{photo}).}
\label{photometry}
\end{centering}
\end{table}

\subsection{The effects of age and metallicity on photometric properties}
\label{eamopp}
This section details how the ages and metallicities derived from 
6dFGS spectra for the central regions of galaxies are related to their
{\it global} photometry.
The analysis considers mainly the youngest (age$<$1.5~Gyr) and oldest
(age$>$10~Gyr) galaxies. By concentrating on the two extremes in age we 
minimise the effects of the age--burst--strength degeneracy on our 
age and metallicity estimates. 
However, it is also important to note that the selection of galaxies with 
extremely young central populations has the effect of biasing the sample 
towards the {\it largest} bursts of star formation.

The following analysis utilises average values of galaxy parameters
binned according to either mass or luminosity.  The derived values are
shown in Table \ref{mol}.  All colours and mass-to-light-ratios for
the binned data were calculated from values in this table. We assume
M$_{B,\odot}$=5.47, M$_{R,\odot}$=4.28 and M$_{K,\odot}$=3.33.

\begin{figure}
\includegraphics[width=6cm,angle=-90]{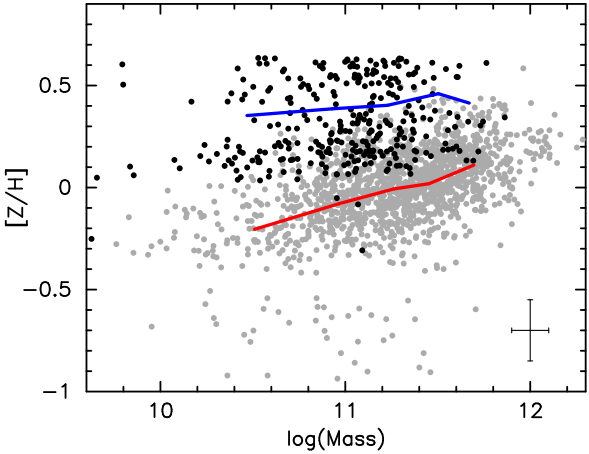}
\caption{The relations between metallicity ([Z/H]) and mass is shown
for young ($<$1.5~Gyr; black symbols) and old ($>$10~Gyr; grey
symbols) galaxies.  Mass-binned average values from Table \ref{mol})
are shown for both old galaxies (red solid line) and young galaxies
(blue solid line). While old galaxies exhibit a clear
mass--metallicity relation, young galaxies do not. }
\label{massmet}
\end{figure}

A mass-metallicity plot of the two sub-samples of our data is shown in
Fig.  \ref{massmet}\footnote{No significant difference was found
between emission-types in these, and subsequent, figures. We therefore
make no distinction between them, although we note that the majority
of young galaxies are also emission galaxies, while, conversely, the
majority of old galaxies are passive galaxies (Fig. \ref{error}).}. In
this plot, galaxies younger than 1.5~Gyr are shown as black
symbols; these galaxies possess an average log(age) of 0.0~dex
(1~Gyr). Galaxies older than 10~Gyr are shown as grey symbols; these
possess an average log(age) of 1.105~dex (12.7~Gyr).  Results of the
mass binning of metallicity values are shown in Fig.  \ref{massmet} as
blue and red lines respectively.  It is evident that the
mass--metallicity relation detected varies with age.  The logarithmic 
slope of the
relation in old galaxies is $\sim$0.25, while the young galaxies are
more metal rich than the old, and the data are consistent with no slope. 
An important aspect of our analysis will
therefore be a consideration of the effects of these trends on
mass-to-light-ratios. We also investigate implications for
optical/near infrared colour-magnitude relations.

\begin{table*}
\begin{centering}
\begin{tabular}{|c|c|c|c|c|c||||||||||||||c|c|c|c|c|c|}  
\hline
\multicolumn{6}{l}{Old, Mass binned}&\multicolumn{6}{l}{Old, Luminosity binned}\\
log(Mass)      &  [Z/H]   &log(age) &M$_B$     &   M$_R$  &  M$_K$  & log(Mass)&  [Z/H]&log(age) & M$_B$  &   M$_R$ &  M$_K$ \\
 \hline				    	 			      	       
11.666          &    0.120 &  1.110 &-20.993   &  -22.162 & -25.497 & 11.755   & 0.137  &  1.107 & -21.311&  -22.507 & -25.870 \\   
11.344          &    0.020 &  1.104 &-20.372   &  -21.506 & -24.769 & 11.411   & 0.038  &  1.101 & -20.564&  -21.715 & -24.983 \\  
11.061          &   -0.040 &  1.109 &-19.796   &  -20.908 & -24.153 & 11.048   &-0.032  &  1.107 & -19.741&  -20.866 & -24.084 \\  
10.778          &   -0.136 &  1.105 &-19.308   &  -20.395 & -23.530 & 10.735   &-0.138  &  1.107 & -19.027&  -20.112 & -23.297 \\  
10.323          &   -0.276 &  1.101 &-18.529   &  -19.546 & -22.623 & 10.330   &-0.213  &  1.104 & -18.232&  -19.222 & -22.340 \\  
\hline								       
\multicolumn{6}{l}{Young, Mass binned}&\multicolumn{6}{l}{Young, Luminosity binned}\\					       
log(Mass)      &  [Z/H]   &log(age) & M$_B$     &   M$_R$  &  M$_K$  &  log(Mass)&  [Z/H]&log(age) & M$_B$   &   M$_R$  &  M$_K$\\         
\hline				 			      	       
11.648          &    0.363 &  0.062  &-21.191   &   -22.216 & -25.557 & 11.430   & 0.317 &   0.042 & -21.366 & -22.382 & -25.728 \\  
11.331          &    0.421 & -0.033  &-20.745   &   -21.733 & -24.959 & 11.209   & 0.311 &  -0.017 & -20.781 & -21.711 & -24.948  \\  
11.062          &    0.394 & -0.054  &-20.432   &   -21.317 & -24.569 & 10.870   & 0.276 &   0.023 & -19.956 & -20.878 & -24.079  \\  
10.781          &    0.381 & -0.011  &-19.804   &   -20.694 & -23.897 & 10.563   & 0.285 &  -0.025 & -19.204 & -20.015 & -23.259  \\  
10.353          &    0.320 & -0.025  &-19.285   &   -20.130 & -23.229 & 10.170   & 0.159 &   0.028 & -18.545 & -19.303 & -22.332 \\  
\hline
\end{tabular}
\caption{Binned average values of mass, metallicity, age and
luminosity of old ($>$10~Gyr) galaxies (top) and young
($<$1.5 Gyr) galaxies (bottom). Binning was carried out in both mass
(left) and K band luminosity (right).}
\label{mol}
\end{centering}
\end{table*}

\subsection{Mass-to-light-ratios in old galaxies}
\label{m2l}
\begin{figure}
\includegraphics[width=8cm,angle=-90]{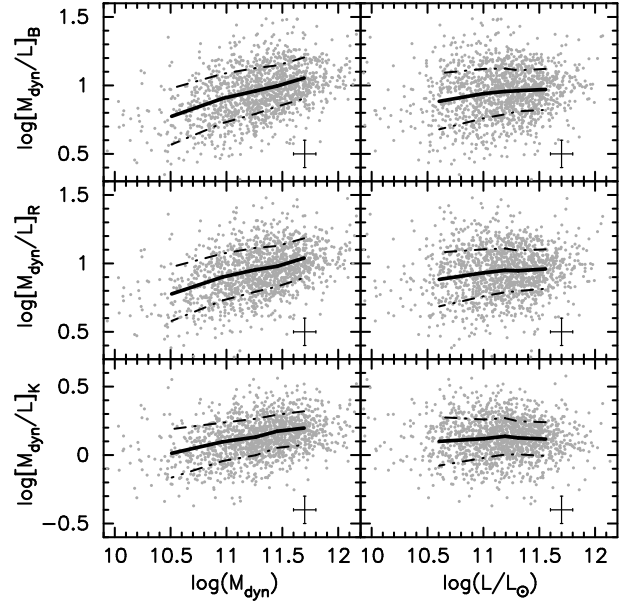}
\caption{The dynamical mass-to-light-ratios of old galaxies in B, R
and K bands are plotted against dynamical mass and K band luminosity.
Solid lines represent averages in five bins along the x-axis. The
extent of the rms scatter is identified by dashed lines. The average
error on individual points shown in the bottom right of each plot.}
\label{m_l1}
\end{figure}

We consider first the $\sim$1500 galaxies identified by our spectral
analysis to possess ages older than 10~Gyr. Plots of [M/L] against
dynamical mass and K band luminosity for these galaxies are shown in
Fig. \ref{m_l1}.  In each plot, the average values of the data in five
bins along the x-axis (Table \ref{mol}) are shown as solid lines. The
extent of the rms scatter in each bin is identified by the dashed
lines.

The observed trends between mass and mass-to-light-ratio in old
 galaxies vary with both waveband and variable against which they are
 plotted. In the K band, we find no trend in mass-to-light-ratio with
 luminosity, i.e.  a slope consistent with zero.  The other five
 plots all show trends with varying, non-zero slope. The logarithmic
 slopes in mass-to-light-ratio with mass are 0.231, 0.214 and 0.157 in
 B, R and K bands respectively, while the logarithmic slopes with
 luminosity are 0.093, 0.080 and 0.022 respectively. The typical
 formal errors on these logarithmic slopes are $\sim$0.01. These
 results are in good agreement with Trujillo, Burkert \& Bell (2004)
 who use a combined SDSS/2MASS catalogue to investigate
 mass-to-light-ratios in the B and K bands (but without the luxury of
 spectroscopy). They found similar slopes in B and K band stellar
 mass-to-light-ratios with luminosity of 0.07 and 0.02~dex
 respectively.

\begin{figure}
\includegraphics[width=8cm,angle=-90]{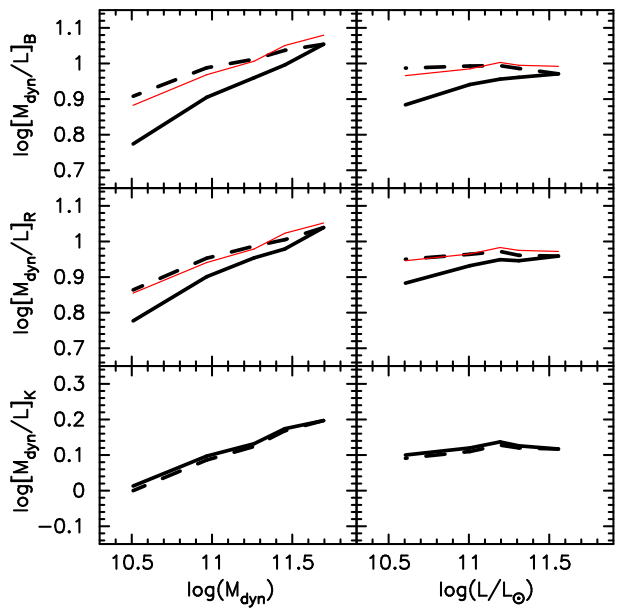}
\caption{Plot of mass-to-light-ratio in B, R and K bands against dynamical 
mass and K band luminosity.  Thick solid lines reproduce the binned
averages of the old galaxies from Fig. \ref{m_l1}. Thick dashed lines
represent these values corrected for metallicity. In B and R band
plots, the K band profile is reproduced (after and offset
$\sim$0.9~dex) as a thin red line. The metallicity corrected values
are remarkably consistent across the three wavebands.}
\label{m_l4}
\end{figure}

While the effects of age are almost entirely eliminated by our
consideration of only old galaxies, the effects of metallicity have
yet to be accounted for.  The effects of the mass--metallicity
relation shown in Fig. \ref{massmet} on mass-to-light-ratios are
investigated in Fig. \ref{m_l4}. This figure again shows the binned
data from Fig. \ref{m_l1} as thick solid lines. Also shown are the
values after correction of the mass-to-light-ratio for differences in
metallicity (dashed lines). The corrections are made by comparing the
luminosities predicted by BC03 models for SSPs of the metallicity (and
age) of each mass or luminosity bin (Table \ref{mol}) with the corresponding
prediction for the highest mass/luminosity bin (Table \ref{mol}).  The
dashed lines in the left hand plots therefore represent the observed
trends between dynamical mass and mass-to-light-ratio in old galaxies (solid
lines) after correction for the mass--metallicity relation evident in
Fig. \ref{massmet} and Table \ref{mol}.  Similarly, the dashed lines
in the right hand plots represent the observed trends between
luminosity and mass-to-light-ratios (solid lines) after correction for the
luminosity--metallicity relation evident Table \ref{mol}.  The dashed
lines can therefore be considered the relations for old galaxies of
fixed metallicity.  We note that the K band
relations are almost unaffected by the corrections. This is due to
the lack of sensitivity of K band luminosity to metallicity (see Table
\ref{photometry}).

The agreement between corrected log[M/L] values is extremely good,
with logarithmic slope against mass of $\sim$0.15 in all three
bands. Against luminosity we find logarithmic slopes in the corrected
values of -0.014, 0.010 and 0.031 in B, R and K bands respectively. 
With formal errors of
$\sim$0.01 these are broadly consistent with a slope of zero. The
remarkable consistency between the results for the three individual
wavebands suggest that the ages, metallicities and photometric 
predictions of the BC03 models are both reliable and internally consistent, 
at least when they are used in the differential sense employed in this work.

The results are also again in good agreement with Trujillo, Burkert \&
Bell (2004), who find slopes equivalent to 0.20 and 0.14 in B and K
bands respectively (compared to our 0.23 and 0.15).  With the benefit
of spectroscopy, we are able to clearly identify metallicity as the
main contributor to the stellar population effects. Our results are
also in good general agreement with the previous studies of both
Bernardi et al. (2003b) and Padmanabhan et al. (2004), who find similar
trends in dynamical mass-to-light-ratios for large numbers of SDSS
galaxies.

\subsection{The fundamental plane}
While our data are not suitable for a direct analysis of the
Fundamental Plane it is nonetheless interesting to compare our results
with such studies in the literature.  To this end we examined the
implications of the varying slopes in mass-to-light-ratio with mass
with waveband on fits to the Fundamental Plane (FP). The implications
can be investigated using the \emph{measured} slopes in log[M/L] with
mass for old galaxies (Table \ref{mol} and Figs \ref{m_l1} and
\ref{m_l4}). Using these values it is possible to predict the FP
(given by log(R)=$\alpha$log($\sigma$)+$\beta$log$<$I$>$+$\gamma$)
that one would expect to find using;

\begin{equation}
   \alpha =\frac{2(1-s)}{1+s},\\
\end{equation}

\noindent and\\

\begin{equation}
   \beta = \frac{-1}{1+s}.\\
\end{equation}

\noindent Where $<$I$>$ is the average surface brightness within the
effective radius (R) and $s$ is the observed slope in log[M/L] with
log(Mass) from Table \ref{mol}.
The virial theorem for homologous systems (i.e. in the absence of a tilt)
would predict $\alpha$ = 2 and $\beta$ = 1.  

The results of the comparisons between predicted $\alpha$ and $\beta$
and the (observational) literature values are shown in Table
\ref{fpcomp}.  The predictions using the observed slopes in
mass-to-light-ratio with mass compare very well to the literature,
values (consistent with the predictions to within $\sim$2$\sigma$). We note
that this is despite both our indirect approach 
as well as literature studies not having the luxury of
isolating old galaxies. 

The comparisons indicate that, while most of the tilt of the FP is 
dynamical in origin, the mass--metallicity relation in 
galaxies (Fig. \ref{massmet}) causes an increasing $\alpha$ and 
$\beta$ with wavelength\footnote {Similar trends in $\alpha$ and $\beta$ with 
wavelength were found by Bernardi et al. (2004) for the galaxy colours in the 
SDSS.} due to the \emph{decreasing}
sensitivity to metallicity of photometry at increasing wavelength. 
This trend culminates in the observed FP tilt in the K band 
being almost identical to the underlying dynamical tilt in the FP 
(see previous section). This is again due to the lack of 
sensitivity of K band luminosity to metallicity in old stellar populations, 
which makes the K band an excellent tool for probing mass-to-light-ratios 
in old (or uniform age) stellar populations.

We have therefore shown our results to be consistent with a broad
range of previous studies of galaxy mass-to-light-ratios and the
Fundamental Plane.  However, the conclusions we draw are not.  To
understand why this is the case, recall that the observed `tilt' in
the Fundamental Plane is often interpreted as a trend in
mass-to-light-ratio with mass or luminosity. Indeed, a trend with mass
of the form [M/L]$\propto$M$^s$ was the starting point of the previous
consideration of the Fundamental Plane. However, careful consideration
of the data show there is a flaw in this interpretation. The
relation [M/L]$\propto$M$^a$ can be simply rearranged into the form
[M/L]$\propto$L$^b$. This gives b=$\frac{a}{1-a}$. Therefore, for
a=0.15, we would expect b=0.18, in stark contrast with the b=0 found.
We must therefore conclude that the simple picture of a trend in
dynamical mass-to-light-ratios with either mass or luminosity is
excluded by our data.\\

In summary, we have shown that our observations are generally in good
agreement with previous literature studies. Furthermore, we have shown
that the mass-metallicity relation with logarithmic slope 0.25 found
in the data is sufficient to explain the differences in the observed
trends in mass-to-light-ratios with mass and luminosity in B, R and K
bands as well as the variations in the apparent `tilt' found in
previous studies of the Fundamental Plane.  However, once this stellar
population effect has been removed, a significant (and consistent)
trend in mass-to-light-ratio with mass remains in all three
wavebands. Finally, we find that the data \emph{rules out} simple
models of variations in dynamical mass-to-light-ratio with mass \emph{or}
luminosity. Further analysis will require a detailed consideration of
the distribution of galaxies in the full 3-dimensional space of the
Fundamental Plane. This is planned for a future paper.  In this work
we proceed by exploring the effects on photometry of young central
spectroscopic ages.

\begin{table*}
\begin{centering}
\begin{tabular}[b]{|c|c|c|c|c|c|c|}
\hline
& \multicolumn{2}{c}{Predictions}&&\multicolumn{2}{c}{Literature}&\\
Band&$\alpha$&$\beta$&&$\alpha$&$\beta$&References\\
\hline
B&1.25&-0.81&&1.20 $\pm$ 0.06 &-0.83 $\pm$ 0.02 &J{\o}rgensen et al. (1996)\\
          &&&&1.33 $\pm$ 0.05 &-0.83 $\pm$ 0.03 &Dressler et al. (1987)\\
\hline
 R&1.29&-0.82&&1.22 $\pm$ 0.09 &-0.84 $\pm$ 0.03 &Colless et al. (2001a)\\
          &&&&1.37 $\pm$ 0.04 &-0.825 $\pm$ 0.01&Gibbons et al. (2001)\\
          &&&&1.38 $\pm$ 0.04 &-0.82 $\pm$ 0.03 &Hudson et al. (1997)\\
\hline
K&1.46&-0.86&&1.53 $\pm$ 0.08 &-0.79 $\pm$ 0.03 &Pahre et al. (1998)\\
\hline
\end{tabular}
\caption{Fundamental Plane coefficients. Values predicted from the
slopes of mass-to-light with mass from Table \ref{mol} (see Section
\ref{m2l}) are shown in the left-hand column.  These are to be compared
with the values from the literature and values derived in this
work. Good agreement is found.}
\label{fpcomp}
\end{centering}
\end{table*}

\subsection{The effects of age on mass-to-light-ratio}
We now turn our attention to the  effects of the young central populations 
detected in our 6dFGS spectra on the global properties of galaxies as 
indicated by their photometry. There are 400 `young' ($<$1.5~Gyr) galaxies 
(see Section \ref{eamopp}) included in this analysis.

\begin{figure}
\center{\includegraphics[width=9cm,angle=-90]{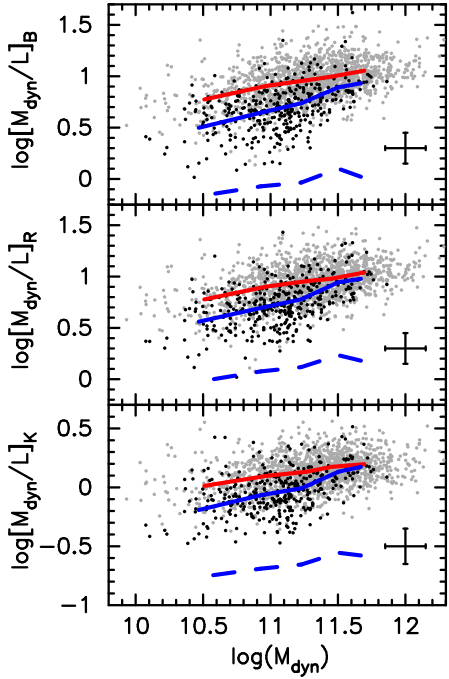}}
\caption{Mass-to-light-ratio in B, R and K bands against mass
for both young and old galaxies. Average errors on data points are
shown in each plot. Mass-binned averages of old and young galaxies are
shown as red and blue lines respectively.  Blue dashed lines represent
the mass-to-light-ratios \emph{expected} in the young galaxies based
on their central spectroscopic ages and metallicities. While young
galaxies deviate from the old population in a sense consistent with
their younger ages, none have mass-to-light-ratios as low as their
central ages would imply.}
\label{m_l2}
\end{figure}

Plots of mass-to-light-ratio with mass for both young and old galaxies
are shown in Fig. \ref{m_l2}. Mass-binned averages of young and old
data are shown as blue and red solid lines respectively.  Also shown
in these plots as dashed blue lines are model predictions of the
effects of the differing ages and metallicities of the young and old
populations in each mass bin. These are based on the comparison of
luminosity values from the BC03 SSP models for the differences in age
and metallicity of the young and old galaxies in each mass bin. A
direct comparison of the data for young galaxies to these lines would
therefore implicitly assume that the young central populations
detected in the spectroscopy pervade the galaxies' populations
\emph{as a whole} - i.e represents the same fraction of the galaxy
population at all radii\footnote{The analysis also implicitly assumes that the
dynamical trends observed in old galaxies, whatever their cause, are
also present in the young galaxies.}.

Young galaxies can be seen to be displaced with respect to the old galaxies in
a sense consistent with their young ages. However, the displacements
are small compared to the predicted values. They also become smaller
towards higher masses. The data therefore show that, on average:
(i) young populations do \emph{not} pervade their host galaxies,
but are instead centrally concentrated; ii) young populations
must constitute relatively small
fractions of total galaxy masses; and (iii) the mass fraction in the
form of a young stellar population must decrease with increasing galaxy
mass.

To quantify these conclusions, for each mass bin, the fraction of
total galaxy stellar mass and luminosity involved in the recent
star-burst (f$_M$ and f$_L$) were estimated.  These estimates were
made by a differential comparison of the BC03 predictions for
mass-to-light-ratios to the observed values. In each waveband, the
difference between the observed mass-to-light-ratios of the young
galaxies from those of the old galaxies in the same mass bin
($\Delta$log[M/L]$_{obs}$) was compared to the difference between
model predictions for young and old populations of appropriate
metallicity ($\Delta$log[M/L]$_{model}$). Denoting these mass-to-light-ratio
differences as s and r respectively, we derive:

\begin{equation}
  \rm   f_M = \frac{10^{-\Delta\log[M/L]_{obs}}-1}{10^{-\Delta\log[M/L]_{model}}-1} \ \ or \ \ \frac{10^{-s}-1}{10^{-r}-1}.
\end{equation}

The derivation of Equation 6 is given in Appendix A.
We estimate the \emph{luminosity} fraction in the recent burst by:

\begin{equation}
\rm f_L = \frac{rf_M}{(1-f_M)+rf_M}.
\end{equation}

Finally, the total mass (M$_{burst}$, in solar masses) is estimated by:

\begin{equation}
\rm M_{burst}  = f_M. M_{gal}.
\end{equation}

Where M$_{gal}$ is the total mass of the galaxy. The derivation of these 
expressions again assumes that the young populations at any 
given mass are seen against a background old population with a metallicity 
and mass-to-light-ratio appropriate to that mass (as given by the mass-binned 
values for old galaxies in Table \ref{mol}). The young stellar population is 
nonetheless assumed to \emph{dominate the central regions} (such that issues 
related to degeneracies with burst-strength do not arise).

\begin{figure}
\center{\includegraphics[width=7cm,angle=-90]{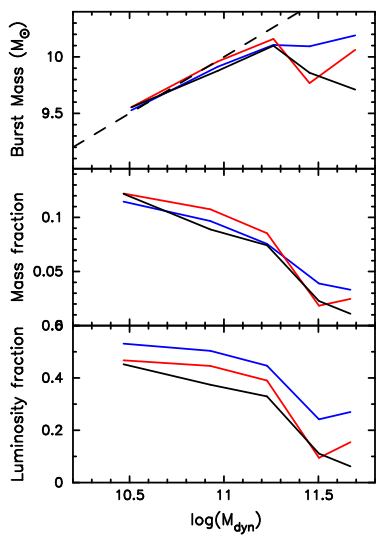}}
\caption{The average luminosity (bottom) and mass (middle) fractions
in recent bursts are plotted against average galaxy mass for the five
mass bins of Figure \ref{m_l1}. Values derived from the B, R and K
bands are shown in blue, red and black lines respectively. Also shown
(top) are estimates of the actual masses (in M$_{\odot}$) involved in
the recent bursts. A fixed 10\% fraction is indicated by the dashed line.
Agreement in mass estimates between wavebands is
extremely good, suggesting an upper limit to the size of star bursts
in these galaxies.}
\label{mass_frac}
\end{figure}

The estimates of mass- and luminosity-fractions and burst-masses are
plotted against galaxy mass in Fig. \ref{mass_frac}. Despite the
(expected) variation in luminosity fractions between wavebands,
agreement between the values for the mass fractions derived from the
three wavebands is extremely good. This suggests that both our
assumption of the same underlying dynamical `tilt' in young galaxies
as exhibited by old galaxies is correct, and that our age and
metallicity corrections are accurate. The trend shows a
decreasing mass fraction with increasing mass falling from $\sim$10\%
at 10$^{10.5}$~M$_{\odot}$ to 2\% at 10$^{11.5}$~M$_{\odot}$. This is
in agreement with the study of Treu et al. (2005), who find the stellar
fraction formed in recent times varies from 20\%-40\% below
10$^{11}$~M$_{\odot}$, to below 1\% above
10$^{11.5}$~M$_{\odot}$.

Also shown in Fig. \ref{mass_frac} are the estimates of average total
mass involved in the recent burst of star formation. Agreement between wavebands is again good. However, the absolute
values given here must be treated with caution, as they are calculated
by the only non-differential application of BC03 models used in this
work. Nevertheless, in a relative sense, the data exhibit a narrow range of
burst-masses and an apparent \emph{upper limit} on their size in high
mass galaxies.

\subsection{Colour-magnitude relations}
As a final consideration we next investigate the implications of our findings 
for colour-magnitude relations. Note that these relations are independent of 
dynamical effects.

\begin{figure}
\includegraphics[width=8cm,angle=-90]{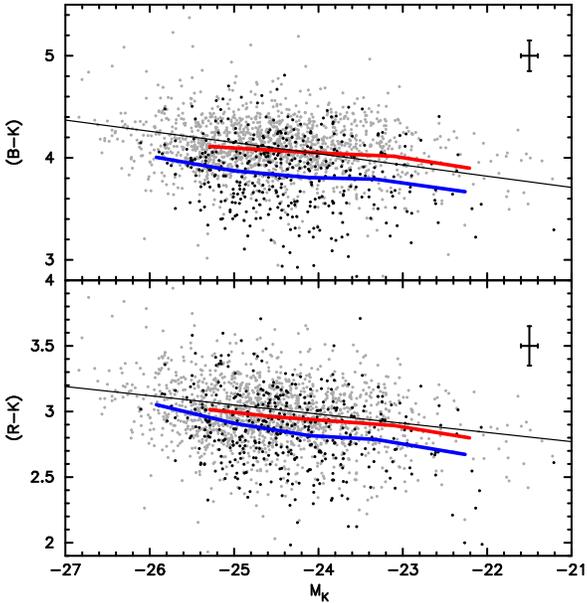}
\caption{The colour-magnitude relations B--K and R--K against M$_K$
are shown for both young and old galaxies. Luminosity-binned average
values are shown as blue and red lines respectively. Young galaxies
are, as expected, bluer than old.}
\label{cmr}
\end{figure}

In Fig. \ref{cmr} we plot the colour-magnitude relations B--K and R--K
against M$_K$.  Galaxies with both old (grey points) and young (black
points) central populations are shown. Averages (binned by luminosity)
of the old and young populations are again shown as solid red and blue
lines respectively. Slopes in the old galaxies are in good accord with
literature values (thin black lines in each plot). As the sample includes 
late-type galaxies, particular amongst the young galaxies, these must
lie close to the `red sequence' in our colour-magnitude
diagrams. This suggests they therefore correspond to the `dusty
red-sequence' of Wolf, Gray \& Meisenheimer (2005). The lack of
galaxies in the `blue-cloud' is almost certainly the result of a
number of selection effects. These include the effects of template
mismatches during velocity dispersion measurements and the removal of
galaxies with poor fits or large errors during age/metallicity
measurement.

Fig. \ref{cmr} shows that, for both colours, young galaxies tend to
be bluer than the old galaxies. However, the offsets are extremely
small. Our results from the previous section indicate that this is
largely the result of the low mass fractions formed in the recent star
formation events.  Fig. \ref{cmr} therefore shows that galaxies lieing
on, or close to, the red sequence may not be `red and dead', but
may still be forming modest numbers of new stars. Indeed, we note that
\emph{most} of the galaxies with young central populations also
exhibit on-going star formation (Fig. \ref{agez}).

\section{Conclusions}
\label{concs}

We have measured ages and metallicities from the 6dFGS spectra of
$\sim$7000 galaxies of all morphological and emission types. We have
demonstrated that the age--metallicity degeneracy has been broken in
this study, particularly with regard to the very young and very old
galaxies used in the subsequent analysis.  These data for the {\it
central regions} are compared with the {\it global} photometry from
2MASS and SuperCOSMOS for the $\sim$6000 galaxies for which such
photometry is available. Combining spectroscopy and photometry we are
able to investigate trends in global B, R and K band
mass-to-light-ratios with the central stellar populations of the
galaxies.

An age-metallicity trend is identified in passive galaxies that is
consistent with the results of similar studies in the literature.  The
stellar populations of galaxies exhibiting emission lines (arising from either
star-bursts or AGN) follow a similar trend, but exhibit younger ages
than their passive counterparts.

To minimise the effects of complex mixtures of stellar populations, we
confine our analysis to the data for very old or very
young galaxies. We find a steep mass-metallicity relation in old
galaxies (logarithmic slope of $\sim$0.25), while in young galaxies
the trend is consistent with zero slope.

Using only the old galaxies, we find a dynamical trend in
mass-to-light-ratios with mass of logarithmic slope $\sim$0.15.
This is in good agreement with values reported in the
literature. However, there is no agreement in the literature as to the
cause of this tilt.  Padmanabhan et al. (2004) concluded that most
of the tilt is due to variations in dark matter content. From
the modelling perspective, Robertson et al. (2006) perform simulations
of a series of mergers and investigate the scaling relations of their
merger remnants. From mergers of gas-rich disk galaxies they derive a
FP with $\alpha$ = 1.55 and $\beta$ = 0.82, in good agreement with
our K band values (see Table 3).  In their models the tilt away from
the virial scaling relation is caused by gas dissipative effects and
not dark matter content.  Trujillo et al. (2004) on the other hand
attribute the tilt to `non-homology', deriving almost identical
contributions to the tilt from dynamical and stellar population
effects as those we find in this work. Our findings, however, also 
show that whatever the cause of the tilt, it is not a simple in
mass-to-light-ratio (or its scatter) with mass or luminosity. Instead
we find a that a more complex description will be required; this will
be a subject of a future paper.

Considering extremely young galaxies, our results show that the
properties of the central stellar populations identified by the
spectroscopy do \emph{not} generally correlate with the global
properties of their host galaxies.  Indeed, the data indicate that
recent bursts of star formation are limited to a mass of
$\lesssim$10$^{10}$~M$_{\odot}$ and are strongly centrally
concentrated. Consequently, in high mass galaxies exhibiting young
central populations, recent star forming events have negligible effect
on their global photometry.  As a consequence we have shown that
galaxies sitting on, or near, the red sequence may not be `red and
dead', but may still be forming modest numbers of new stars.

We therefore conclude that the young ages found in studies of galaxy
centres in this and other spectroscopic studies should not be
considered representative of the age of their host galaxies as a
whole. They may, however, be reasonably interpreted as indicators of
recent assembly events.  As a result of these considerations, the
small scatter in photometric scaling relations, which initially
motivated the monolithic collapse models, ceases to be inconsistent
with either the hierarchical merging model or the results of the Lick
index studies indicating young central populations.

In future work we will include the results from the completed 6dF
Galaxy Survey and report on the full chemical properties (including
$\alpha$-abundance ratios) of the central regions of galaxies. The
effective radii, peculiar velocities and environmental parameters of
the 6dFGS sample, currently being analysed by the 6dFGS team,
will be included. This will allow not only more accurate determination
of photometric properties and a more detailed analysis of the
Fundamental Plane, but also the investigation of the strong {\it
quantitative} predictions of trends in galaxy formation/assembly
epochs and metals-content with galaxy mass and environment.\\

\noindent{\bf Acknowledgements}

This publication makes use of data products from the Two Micron All
Sky Survey (2MASS), which is a joint project of the University of
Massachusetts and the Infrared Processing and Analysis
Center/California Institute of Technology, funded by the National
Aeronautics and Space Administration and the National Science
Foundation.

The authors acknowledge the data analysis facilities provided by IRAF, which
is distributed by the National Optical Astronomy Observatories and
operated by AURA, Inc., under cooperative agreement with the National
Science Foundation. We thank the Australian Research Council for funding 
that supported this work. 

The measuring of line values from 6dFGS spectra was begun in Philip
Lah's Honours Thesis 2003 ``Exploring the Stellar Population of
Early-Type Galaxies in the 6dF Galaxy Survey'' which looked at the
6dFGS Early Data Release, (December 2002).  This Honours thesis was
supervised by Matthew Colless and Heath Jones.  Assistance with the
software used in this work was provided by Craig Harrison.  Lachlan
Campbell provided additional assistance with the 6dFGS data
particularly with the velocity dispersions and 6dF arc spectra. The
authors also thank Chris Blake for valuable input.

\begin{appendix}
\label{app}

\subsection{Appendix 1. Mass fraction derivation}
\label{mfder}

The logarithm of the mass-to-light ratio in a galaxy comprising two populations; one young and one old, can be written:

\begin{equation}
\rm log[M/L]_{obs}=log[\frac{M}{L_{old}+L_{young}}].
\end{equation}

This parameterisation assumes that the mass of the galaxy is unaffected by the 
starburst that generates the young stellar population (the galaxy converts 
some of its own mass into stars).

Now, expressing the fraction of the observed stellar mass in young stars as 
f$_M$ and the ratio of mass-to-light-ratios of young and old populations as r, 
we can write:

\begin{equation}
\rm log[M/L]_{obs}=log[\frac{M}{(1-f_M)L_{old}+rf_ML_{old}}].
\end{equation}

\noindent Rearranging:

\begin{equation}
\rm log[M/L]_{obs}=log[M/L]_{old}-log((1-f_M)+rf_M).
\end{equation}

\noindent Defining:

\begin{equation}
\rm \Delta~log[M/L]_{obs}=log[M/L]_{obs}-log[M/L]_{old}.
\end{equation}

\noindent We get;

\begin{equation}
\rm \Delta~log[M/L]_{obs}=-log((r-1)f_M+1).
\end{equation}

\noindent Rearranging:

\begin{equation}
\rm 10^{-\Delta~log[M/L]_{obs}}=(r-1)f_M+1,
\end{equation}

and;

\begin{equation}
\rm f_M=\frac{10^{-\Delta~log[M/L]_{obs}}-1}{r-1}
\end{equation}

\noindent The ratio of mass-to-light-ratios (r), as defined above can be 
estimated using the BC03 models of appropriate age and metallicity: 

\begin{equation}
\rm r=10^{log[M/L]_{BC03 old}-[log[M/L]_{BC03 young}}
\end{equation}

\noindent I.e:

\begin{equation}
\rm r= 10^{-\Delta\log[M/L]_{model}}
\end{equation}

\noindent We thus derive the fraction of the stellar mass in the 
1~Gyr old population as:

\begin{equation}
  \rm   f_M = \frac{10^{-\Delta\log[M/L]_{obs}}-1}{10^{-\Delta\log[M/L]_{model}}-1}
\end{equation}

We note that, by the differential use of both the observations and 
models, mass fractions are derived \emph{without} knowledge of the true
stellar to dynamical mass ratios.\\

\end{appendix}

\noindent{\bf References}\\
\noindent
Baldry I.K. et al., 2002, ApJ, 569, 582\\ 
Baum W.A., 1959, IAU Symp. 10, The Hertzsprung-Russell Diagram, ed. J. L. Greenstein (Paris: IAU), 23\\
Bernardi M. et al., 2003a, AJ, 125, 1882\\
Bernardi M. et al., 2003b, AJ, 125, 1866\\
Bernardi M. et al., 2006, AJ, 131, 2018\\
Bernardi M., 2007, AJ, 133, 1954\\
Blumenthal G.R., Faber S.M., Primack J.R., Rees M.J., 1984, Nature, 311, 517\\
Bower R.G., Lucy J.R., Ellis R.S., 1992a,  MNRAS, 254, 589\\
Bower R.G., Lucy J.R., Ellis R.S., 1992b,  MNRAS, 254, 601\\
Bruzual A.G., Charlot S., 2003, MNRAS, 344, 1000\\
Caldwell N., Rose J.A., Concannon K.D., 2003, AJ, 125, 2891\\
Cappellari M. et al., 2006, MNRAS, 366, 1126\\
Carlberg R.G., 1984, ApJ, 286, 403\\
Chang R., Gallazzi A., Kauffmann G., Charlot S., Ivezi\'{c} \~{Z}., Brinchmann J., Heckman T.M., 2006, MNRAS, 366, 717\\
Cid Fernandes R., Gu Q., Melnick J., Terlevich E., Terlevich R., Kunth D., Rodrigues Lacerda R., Joguet B., 2004, MNRAS, 355, 273\\
Cid Fernandes R, Mateus, A., Sodr\'{e} L., Stasinska G., Gomes J.M., 2005, MNRAS, 358, 363\\
Clemens M.S., Bressan A., Nikolic B., Alexander P., Annibali F., Rampazzo R., 2006, MNRAS, 370, 702\\
Colless M., Saglia R.P., Burstein D., Davies R.L., McMahan Jr R.K., Wegner G., 2001a, MNRAS, 321, 277\\
Colless M. et al., 2001b, MNRAS, 328, 1039\\ 
Collobert M., Sarzi M., Davies R.L., Kuntschner H., Colless M., 2006, MNRAS, 370, 1213\\
De Lucia G., Springel V., White S.D.M., Croton D., Kauffmann G., 2006, MNRAS, 366, 499\\
Dressler A., Lynden-Ball D., Burstein D., Faber S.M., Terlevich R., Wegner G., 1987, ApJ, 313, 42\\
Eisenstein D.J. et al., 2003, ApJ, 585, 694\\
Fernandes R.C., Mateus, A., Sodr\'{e} L., Stasinska G., Gomes J.M., 2005, MNRAS, 358, 363\\
Gallazzi A., Charlot S., Brinchmann J., White S.D.M., Tremonti C., 2005, MNRAS, 362, 41\\
Gebhardt, K. et al., ApJ, 539, L13\\
Gibbons R.A., Fruchter A.S., Bothun G.D., 2001, AJ, 121, 649\\
Gonz\'{a}lez J. J., 1993, PhD thesis, Univ. California\\
Hambly N.C., Irwin M.J., MacGillivray H.T., 2001, MNRAS, 326, 1295\\
Hudson M.J., Lucey J.R., Smith R.J., Steel J., 1997, MNRAS, 291, 488\\
Jarrett T.H., Chester T., Cutri R., Schneider S., Skrutskie M., Huchra J.P.,  2000a, AJ, 119, 2498\\
Jarrett T.H., Cheste, T., Cutri R., Schneider S., Rosenberg J., Huchra J.P., Mader J., 2000b, AJ, 120, 298\\
Jones D.H. et al., 2004, MNRAS, 355,747\\
Jones D.H., Peterson B.A., Colless M., Saunders W.,  2006, MNRAS, 369, 25\\
J{\o}rgensen I., Franx M., Kj{\ae}rgaard P., 1995, MNRAS, 276, 1341\\
J{\o}rgensen I., Franx M., Kj{\ae}rgaard P., 1996, MNRAS, 280, 167\\
Kauffmann G., 1996, MNRAS, 281, 487\\
Kauffmann G. et al., 2003, MNRAS, 346, 1055\\
Kauffmann G. et al., 2003a, MNRAS, 341, 33\\
Kauffmann G. et al., 2003b, MNRAS, 341, 54\\
Kewley L.J., Dopita M.A., Sutherland R.S., Heisler C.A., Trevena J., 2001, ApJ, 556, 121\\
Korn A.J., Maraston C., Thomas D., 2005, A\&A, 438, 685\\
Larson R.B., 1974, MNRAS, 166, 585\\
Lewis I. et al., 2002, MNRAS, 334, 673\\
Mehlert D., Thomas D., Saglia R.P., Bender R., Wegner G., 2003, A\&A, 407, 423\\
Moore S.A.W., 2001, Ph.D thesis, Univ of Durham\\
Padmanabhan N. et al., 2004, New Astron., 9, 329\\
Pahre M.A., Djorgovski S.G., de Carvalho R.R., 1998, AJ, 116, 1591\\
Proctor R.N., Sansom A.E., 2002, MNRAS, 333, 517\\
Proctor R.N., Forbes D.A., Beasley M.A.,  2004a, MNRAS, 355, 1327\\
Proctor R.N., Forbes D.A., Hau G.K.T., Beasley M.A., De Silva G.M., Contreras R., Terlevich A.I., 2004b, MNRAS, 349, 1381\\
Proctor R.N., Forbes D.A., Forestell A., Gebhardt K., 2005, MNRAS, 362, 857\\
Robertson B., Cox T.J., Hernquist L., Franx M., Hopkins P.F., Martini P., Springel V., 2006, ApJ, 641, 21\\
Smith R.J., Hudson M.J., Lucey, J.R., Nelan, J.E., Wegner, G.A., 2006, MNRAS, 369, 1419\\
Spergel D.N. et al., 2003, ApJS, 148, 175\\
Terlevich A.I., Forbes D.A., 2002, MNRAS, 330, 547\\
Thomas D., Maraston C., Bender R., 2003, MNRAS, 339, 897\\
Thomas D., Maraston C., Bender R., Mendes de Oliveira C., 2005, ApJ, 621, 673\\
Thomas D., Maraston C., Korn A., 2004, MNRAS, 351, 19\\
Trager, S., Faber, S., Worthey, G., Gonzalez, J., 2000, AJ, 120, 165\\
Tremonti C. et al., ApJ, 613, 898\\
Treu T., Ellis R.S., Liao T.X., van Dokkum P.G., Tozzi P., Coli, A., Newman J., Cooper M.C., Davis M., 2005, ApJ, 633, 174\\
Trujjillo I., Burkert A., Bell E.F., 2004, AJ, 600, L39\\
White S.D.M., Rees M.J., 1978, MNRAS, 183, 341\\
Wolf C., Gray M.E., Meisenheimer K., 2005, A\&A, 443, 435\\
Worthey G., 1994, ApJS, 95,107\\
Worthey G., Ottaviani D.L., 1997, ApJS, 111, 377\\
York D.G. et al., 2000, AJ, 120, 1579\\

\label{lastpage}
\end{document}